\documentclass[traditabstract]{aa} 
\usepackage{graphicx,color}
\begin{document}
   \title{Growth of perturbations in an expanding universe with
Bose-Einstein
condensate dark matter}


   \author{P.H. Chavanis
          }

   \institute{Laboratoire de Physique Th\'eorique (IRSAMC), CNRS and UPS, Universit\'e de Toulouse, France
             \email{chavanis@irsamc.ups-tlse.fr}
             }



  \abstract {We study the growth of perturbations in an expanding
  Newtonian universe with Bose-Einstein condensate dark matter. We
  first ignore special relativistic effects and derive a differential
  equation governing the evolution of the density contrast in the
  linear regime taking into account quantum pressure and
  self-interaction. This equation can be solved analytically in
  several cases. We argue that an attractive self-interaction can
  enhance the Jeans instability and fasten the formation of
  structures. Then, we take into account pressure effects (coming from
  special relativity) in the evolution of the cosmic fluid and add the
  contribution of radiation, baryons and dark energy (cosmological
  constant). For a BEC dark matter with repulsive self-interaction
  (positive pressure) the scale factor increases more rapidly than in
  the standard $\Lambda$CDM model where dark matter is pressureless
  while for a BEC dark matter with attractive self-interaction
  (negative pressure) it increases less rapidly. We study the linear
  development of the perturbations in these two cases and show that
  the perturbations grow faster in a BEC dark matter than in a
  pressureless dark matter. This confirms a recent result of Harko
  (2011). Finally, we consider a ``dark fluid'' with a generalized
  equation of state $p=(\alpha \rho+k\rho^2)c^2$ having a component
  $p=k\rho^2 c^2$ similar to a BEC dark matter and a component
  $p=\alpha\rho c^2$ mimicking the effect of the cosmological constant
  (dark energy). We find optimal parameters that give a good agreement
  with the standard $\Lambda$CDM model assuming a finite cosmological
  constant.}


   \maketitle
%

\section{Introduction}

Several astrophysical observations of distant type Ia supernovae have revealed that the content of the universe is made of about $70\%$ of dark energy, $25\%$ of dark matter and $5\%$ of baryonic (visible) matter (Riess et al. 1998, Perlmutter et al. 1999, de Bernardis et al. 2000, Hanany et al. 2000).  Thus, the overwhelming preponderance of matter and energy in the universe is believed to be dark i.e. unobservable by telescopes.  The dark energy is responsible for the accelerated expansion of the universe. Its origin is mysterious and presumably related to the cosmological constant. Dark energy is usually interpreted as a vacuum energy and it behaves like a fluid with negative pressure. Dark matter also is mysterious. The suggestion that dark matter may constitute a large part of the universe was raised by Zwicky in 1937. He realized that some mass was ``missing'' in order to account for observations. This missing mass problem was confirmed later by more accurate measurements (Borriello \& Salucci 2001). The rotation curves of neutral hydrogen clouds in spiral galaxies measured from the Doppler effect are found to be roughly flat with a typical rotational velocity $v_\infty\sim 200\, {\rm km/s}$ up to the   maximum observed radius of about $50$ kpc. This mass profile is much more extended than the distribution of starlight which typically converges within $\sim 10$ kpc. This implies that galaxies are surrounded by an extended halo of dark matter whose mass $M(r)\sim r v_\infty^2/G$ increases linearly with radius. 
Although some authors like Milgrom (1983)  propose a modification of Newton's law (MOND theory) to explain the rotation curves of spiral galaxies without invoking dark matter, the dark matter hypothesis is favored by most astrophysicists.

The nature of dark matter (DM) is one of the most important puzzles in modern physics and cosmology. A wide ``zoology'' of exotic particles that could form dark matter has been proposed. In particular, many grand unified theories in particle physics predict the existence of various exotic  bosons (e.g. axions, scalar neutrinos, neutralinos) that should be present in considerable abundance in the universe and comprise (part of) the cosmological missing mass (Primack et al. 1998,  Overduin \& Wesson 2004). Even if the bosonic particles have never been detected in accelerator experiments, they are considered as leading candidates of dark matter and might play a significant role in the evolution and in the structure of the universe.

If dark matter is made of bosons, they should have formed compact gravitating  Bose-Einstein condensates (BEC) such as boson stars. Boson stars were introduced by Kaup (1968)  and Ruffini \& Bonazzola  (1969) in the sixties. Early  works on boson stars  (Thirring 1983,   Breit et al. 1984,  Takasugi \& Yoshimura 1984, van der Bij \& Gleiser 1987) were motivated by the axion field that was proposed as a possible solution to the strong CP problem in QCD. For particles with mass $m\sim 1\, {\rm GeV/c^2}$, the maximum mass of a boson star, called the Kaup mass, is much smaller than the solar mass ($M_{Kaup}\sim 10^{-19}M_\odot$!) so that these {\it mini boson stars}, like axion black holes, are not very  astrophysically relevant. Some authors (Baldeschi et al. 1983, Sin 1994, Hu et al. 2000) have proposed that dark matter halos could be giant systems of ``Bose liquid'' but in that case the mass of the bosons must be extremely small ($m\sim 10^{-24}\, {\rm eV}$) to yield masses consistent with the mass of galactic halos. Such an ultralight scalar field was called ``fuzzy cold dark matter'' (FCDM) by Hu  et al.  (2000) who discussed its overall cosmological behavior. On the other hand, Colpi et al. (1986) have shown that if the bosons have a self-interaction, then the mass of the boson stars can considerably increase, even for a small self-interaction. For $m\sim 1\, {\rm GeV/c^2}$ it becomes of the order of the solar mass so that dark matter could be made of numerous boson stars. On the other hand, for $m\sim 1\, {\rm eV/c^2}$, the mass of the boson stars becomes of the order of the mass of the galactic halos. Therefore, some authors (Lee \& Koh 1996, Peebles 2000, Goodman 2000, Arbey et al. 2003, B\"ohmer \& Harko 2007) proposed that dark matter halos themselves could be in the form of gigantic self-gravitating Bose-Einstein condensates with short-range interactions described by a single wave function $\psi({\bf r},t)$. In the Newtonian limit, which is relevant at the galactic scale, the evolution of this wave function  is governed by the Gross-Pitaevskii-Poisson (GPP) system  (B\"ohmer \& Harko 2007). Using the Madelung (1927) transformation, the GP equation turns out to be equivalent to hydrodynamic (Euler) equations involving an isotropic classical pressure due to short-range interactions (scattering) and an anisotropic  quantum pressure (or a quantum potential) arising from the Heisenberg uncertainty principle. For a standard BEC with quartic self-interaction, the equation of state is that of a polytrope with index $n=1$. At large scales, scattering and quantum effects are negligible and one recovers the classical hydrodynamic equations of cold dark matter (CDM) models which are remarkably successful in explaining the large-scale structure of the universe. At small scales, gravitational collapse is prevented by the (repulsive) scattering or by the uncertainty principle. This may be a way to solve the problems of the CDM model such as the cusp problem and the missing satellite problem (Hu et al. 2000).

In our previous papers (Chavanis 2011, Chavanis \& Delfini 2011), we performed an exhaustive study of the equilibrium configurations of a Newtonian self-gravitating BEC with short-range interactions. For a given value of the scattering length, we obtained the mass-radius relation $M(R)$ connecting the non-interacting case (corresponding to small masses) studied by Ruffini \& Bonazzola (1969) to the Thomas-Fermi (TF) limit (corresponding to large masses) investigated by B\"ohmer \& Harko (2007). We also considered the case of attractive self-interaction. This corresponds to a  negative scattering length ($a_s<0$) yielding a negative pressure.  In that case, we found the existence of a maximum mass $M_{max}=1.012\hbar/\sqrt{|a_s|Gm}$ above which the system becomes unstable. It turns out that this mass is ridiculously small (being possibly as small as the Planck mass $M_P=2.18\, 10^{-8} {\rm kg}!$)  meaning that a self-gravitating BEC with attractive short-range interactions is extremely  unstable (except if the mass of the bosons is extraordinarily small like in Hu et al. 2000). We proposed that the scattering length could be negative in the early universe and that it could help for the formation of structures\footnote{This is a potentially interesting idea because, in an expanding universe, the condensation process is often regarded to be too slow to account for the formation of structures (Bonnor 1957, Harrison 1967). An attractive self-interaction could substantially  enhance the instability.}.  Then, the scattering length could become positive and help to stabilize the structures against gravitational collapse. Of course, the mechanism by which the scattering length  changes sign remains to be established so that this idea is highly speculative. We may note that some atoms in terrestrial BEC experiments are reported to have negative scattering lengths (Dalfovo et al. 1999). Therefore, the possibility of negative pressure for a BEC can be contemplated. Furthermore, it has been experimentally demonstrated that, under certain conditions, it is possible  to manipulate the sign and value of the scattering length (Fedichev et al. 1996). The extension of these ideas to cosmology remains an (important) problem that we shall not discuss further here. {\it We shall assume that the dark matter in the universe is a BEC with  repulsive ($a_s>0$) or attractive ($a_s<0$) self-interaction and theoretically explore the consequences of this hypothesis.}

If dark matter halos are BECs, they have probably formed by Jeans instability. The gravitational instability  of a scalar field equivalent  to a BEC was considered by Khlopov et al. (1985), Bianchi {et al} (1990), Hu et al. (2000)  and Sikivie \& Yang (2009). However, these authors started their analysis from relativistic field equations and did not take into account the self-interaction of the particles. In our previous paper (Chavanis 2011), we studied the Jeans  instability  of a self-gravitating BEC with short-range interactions described by the Gross-Pitaevskii-Poisson system in a static universe. We considered both attractive and repulsive self-interaction and found  that when the scattering length is negative the growth rate of the instability increases with respect to the case where the scattering length is positive or zero. The next step is to study the Jeans instability of a BEC in an expanding universe. This is the topic of the present paper.

In the first part of the paper (Secs. \ref{sec_gpp}-\ref{sec_eds}), we neglect relativistic effects and  use the equations of Newtonian cosmology introduced by Milne (1934) and McCrea \& Milne (1934). We study the linear development of perturbations of a self-gravitating BEC with short-range interactions in an expanding Einstein-de Sitter (EdS) universe. In the TF approximation, the equation of state of a BEC with quartic self-interaction is that of a polytrope of index $\gamma=2$ ($n=1$). In the non-interacting case, we find that the BEC behaves similarly to a polytrope of index $\gamma=5/3$ ($n=3/2$) but with a quantum mechanically modified Jeans length. In the two cases, the equation for the density contrast can be solved analytically in terms of Bessel functions.

In the second part of the paper (Sec. \ref{sec_pressure}), we take
special relativistic effects into account and use the equations of
Newtonian cosmology with pressure introduced by McCrea (1951). In that
case, the evolution of the cosmic fluid depends on the equation of
state. We consider a BEC dark matter with equation of state $p=k\rho
c^2$ and add the contribution of radiation, baryons and dark energy
(cosmological constant). For a BEC dark matter with repulsive
self-interaction $k>0$ (positive pressure) the scale factor increases
more rapidly than in the standard $\Lambda$CDM model where dark matter
is pressureless ($k=0$) while for a BEC dark matter with attractive
self-interaction $k<0$ (negative pressure) it increases less rapidly.
We study the linear development of the perturbations in these two
cases and show that the perturbations grow faster in a BEC dark matter
than in a pressureless dark matter.

In the third part of the paper (Sec. \ref{sec_dark}), we consider a
``dark fluid'' with a generalized equation of state $p=(\alpha
\rho+k\rho^2)c^2$ having a component $p=k\rho^2 c^2$ similar to a BEC
dark matter and a component $p=\alpha\rho c^2$ mimicking the effect of
the cosmological constant (dark energy). We find optimal parameters
that give a good agreement with the standard $\Lambda$CDM model.

While our series of papers on this subject was in course of redaction,
Harko (2011) published a paper where he also considered the formation
of structures in an expanding universe made of BEC dark matter. Our
contribution is complementary to Harko's work and confirms his main
results. In addition, we address the following issues: (i) we study
the effect of the quantum pressure in the first part of the paper;
(ii) we use different relativistic hydrodynamic equations to model the
cosmic fluid in the second part of the paper; (iii) we study a
generalized equation of state in the third part of the paper; (iv) we
consider both positive and negative scattering lengths throughout the
paper. The papers of B\"ohmer \& Harko (2007) and Harko (2011) show
that a cosmic BEC can be an interesting model for the dark matter of
the universe. Our series of papers, which develop this idea, go in the
same direction.

Finally, in a different perspective, Widrow \& Kaiser (1993) have proposed to describe a classical collisionless self-gravitating system by the  Schr\"odinger-Poisson system. In this approach, the constant $\hbar$ is not the Planck constant, but rather an adjustable parameter that controls the spatial resolution $\lambda_{deB}$ through a de Broglie relation $\lambda_{deB}=\hbar/mv$. It is argued that when $\hbar\rightarrow 0$, the Vlasov-Poisson system is recovered and that a finite value of $\hbar$ provides a small-scale regularization of the dynamics. In that case, the Schr\"odinger-Poisson system has nothing to do with quantum mechanics since it aims at describing the evolution of classical collisionless matter under the influence of gravity (in static or expanding universes). Still, on a mathematical point of view, these equations are equivalent to those describing self-gravitating BECs without self-interaction. Therefore, the results of the first part of our paper can have application  in that context, independently of quantum mechanics.

\section{The Gross-Pitaevskii-Poisson system}
\label{sec_gpp}

Following B\"ohmer \& Harko (2007), we assume that dark matter is a Bose-Einstein condensate. A self-gravitating BEC with short-range interactions is described by the  Gross-Pitaevskii-Poisson system
\begin{equation}
\label{gpp1}
i\hbar \frac{\partial\psi}{\partial t}=-\frac{\hbar^2}{2m}\Delta\psi+m(\Phi+h(\rho))\psi,
\end{equation}
\begin{equation}
\label{gpp2}
\Delta\Phi=4\pi G Nm |\psi|^2-\Lambda,
\end{equation}
where $\rho=Nm|\psi|^2$ is the density, $\Phi$ the gravitational
potential and $h(\rho)=\int u_{SR}({\bf r}-{\bf r}')\rho({\bf r}',t)\,
d{\bf r}'$ an effective potential taking into account small-scale
interactions. For the sake of generality, we have included the
cosmological constant $\Lambda$ in the Poisson equation\footnote{We
may note that the cosmological constant $\Lambda$ plays the same role
as a global rotation $\Omega$ of the universe if $\Phi$ is interpreted
as an effective gravitational potential
$\Phi_{eff}=\Phi-\frac{1}{2}({\bf \Omega}\times {\bf r})^2$ in the
rotating frame. Indeed, the Poisson equation becomes
$\Delta\Phi_{eff}=4\pi G \rho-2\Omega^2$ which allows the
identification $\Lambda=2\Omega^2$.}. We write the wave function in
the form $\psi({\bf r},t)=A({\bf r},t)e^{iS({\bf r},t)/\hbar}$ where
$A$ and $S$ are real, and make the Madelung (1927) transformation
\begin{equation}
\label{gpp3}
\rho=Nm|\psi|^2=NmA^2, \qquad  {\bf u}=\frac{1}{m}\nabla S,
\end{equation}
where $\rho({\bf r},t)$ is the density field and ${\bf u}({\bf r},t)$ the velocity field. We note that the flow is irrotational since $\nabla\times {\bf u}={\bf 0}$.  With this transformation, it can be shown that the Gross-Pitaevskii equation (\ref{gpp1}) is equivalent to the barotropic Euler equations with an additional term $Q$ called the quantum potential (or quantum pressure). Indeed, one obtains the set of equations
\begin{equation}
\label{gpp4}
\frac{\partial\rho}{\partial t}+\nabla\cdot (\rho {\bf u})=0,
\end{equation}
\begin{equation}
\label{gpp5}
\frac{\partial {\bf u}}{\partial t}+({\bf u}\cdot \nabla){\bf u}=-\frac{1}{\rho}\nabla p-\nabla\Phi-\frac{1}{m}\nabla Q,
\end{equation}
\begin{equation}
\label{gpp6}
\Delta\Phi=4\pi G \rho-\Lambda,
\end{equation}
with
\begin{equation}
\label{gpp7}
Q=-\frac{\hbar^2}{2m}\frac{\Delta \sqrt{\rho}}{\sqrt{\rho}}.
\end{equation}
The pressure $p(\rho)$ is determined by the effective potential $h(\rho)$, playing the role of an enthalpy (Chavanis 2011), through the relation $p'(\rho)=\rho h'(\rho)$. For a contact pair interaction $u_{SR}({\bf r}-{\bf r}')=g\delta({\bf r}-{\bf r}')$, the effective potential $h(\rho)=g\rho$ where $g={4\pi a_s\hbar^2}/{m^3}$ is the pseudo-potential and  $a_s$ is the s-scattering length (Dalfovo et al. 1999). For the sake of generality, we allow $a_s$ to be positive or negative. The corresponding equation of state is
\begin{equation}
\label{gpp8}
p=\frac{2\pi a_s\hbar^2}{m^3}\rho^{2}.
\end{equation}
It corresponds to a polytropic equation of state of the form
\begin{equation}
\label{gpp9}
p=K\rho^{\gamma},\qquad \gamma=1+\frac{1}{n},
\end{equation}
with polytropic index $n=1$ (i.e. $\gamma=2$) and polytropic constant $K={2\pi a_s\hbar^2}/{m^3}$. 

The effective potential corresponding to the polytropic equation of state (\ref{gpp9}) is
\begin{equation}
\label{gpp10}
h(\rho)=\frac{K\gamma}{\gamma-1}\rho^{\gamma-1}.
\end{equation}
This leads to a GP equation of the form
\begin{equation}
\label{gpp11}
i\hbar \frac{\partial\psi}{\partial t}=-\frac{\hbar^2}{2m}\Delta\psi+m(\Phi+\kappa |\psi|^{2/n})\psi,
\end{equation}
where $\kappa=K(n+1)(Nm)^{1/n}$. This is the usual form of the GP
equation considered in the literature (Sulem \& Sulem 1999). The
standard BEC, with a quartic interaction $p\propto \rho^2\propto
|\psi|^4$, corresponds to $n=1$.  We may also recall that classical
and ultra-relativistic fermion stars are equivalent to polytropes with
index $n=3/2$ and $n=3$ (Chandrasekhar 1939). The GPP system
(\ref{gpp11})-(\ref{gpp2}) with $n=3/2$ has been studied by Bilic {et
al.} (2001) in relation to the formation of white dwarf stars by
gravitational collapse. They showed that the quantum pressure can
regularize the dynamics at small-scales.

For an isothermal equation of state
\begin{equation}
\label{gpp12}
p=\rho \frac{k_B T_{eff}}{m},
\end{equation}
the effective potential is
\begin{equation}
\label{gpp13}
h(\rho)=\frac{k_B T_{eff}}{m}\ln\rho,
\end{equation}
and the GP equation reads
\begin{equation}
\label{gpp14}
i\hbar \frac{\partial\psi}{\partial t}=-\frac{\hbar^2}{2m}\Delta\psi+(m\Phi+{2k_B T_{eff}}\ln|\psi|)\psi.
\end{equation}
Interestingly, we note that a nonlinear Schr\"odinger equation with a logarithmic potential similar to Eq. (\ref{gpp14}) has been introduced long ago by Bialynicki \& Mycielski (1976) as a possible generalization of the Schr\"odinger equation in quantum mechanics.

{\it Remark:} for a BEC at $T=0$, the pressure arising in the Euler equation (\ref{gpp5}) has a meaning different from the kinetic pressure of a normal fluid at finite temperature. It is due to the self-interaction of the particles encapsulated in the effective potential $h(\rho)$ and not to thermal motion (since $T=0$). As a result, the pressure can be {\it negative} (!) contrary to a kinetic pressure. This is the case, in particular, for a BEC described by the equation of state (\ref{gpp8}) when the scattering length $a_s$ is negative.  In terrestrial BEC experiments, some atoms like $^7{\rm Li}$ have a negative scattering length (Fedichev et al. 1996).  On the other hand, the constant $T_{eff}$ appearing in Eq. (\ref{gpp12}) is just an ``effective'' temperature since it arises from a particular form of self-interaction and has nothing to do with the kinetic temperature (which here is $T=0$). In particular, this effective temperature can be negative. In that respect, we note that linear equations of state $p=\alpha\rho c^2$ with $\alpha<0$ have been introduced heuristically to account for the accelerated expansion of the universe (see the review of Peebles \& Ratra 2003 and Sec. \ref{sec_dark}).

\section{Newtonian cosmology}
\label{sec_nc}

In this first part of the paper, we use the Newtonian cosmology introduced by Milne and McCrea \& Milne in 1934. The great advantage of the Newtonian treatment is its simplicity\footnote{It is surprising to realize that Newtonian cosmology was developed after, and was influenced by,  the cosmological models based on Einstein's theory of general relativity. It could have been developed much earlier. As Milne writes: ``It seems to have escaped previous notice that whereas the theory of the expanding universe is generally held to be one of the fruits of the theory of relativity, actually all the phenomena observable at present could have been predicted by the founders of mathematical hydrodynamics in the eighteenth century, or even by Newton himself''.}. Furthermore, the equations are identical with those derived using the theory of general relativity, provided the pressure is negligible in comparison with the energy density $\rho c^2$ where $c$ is the speed of light (we shall go beyond this limitation in the second part of the paper). This makes this approach attractive. Furthermore, its simplicity allows us to introduce novel ingredients such as the quantum pressure which is derived from the classical Gross-Pitaevskii equation. The validity and limitation of Newtonian cosmology have been discussed by various authors such as Layzer (1954), McCrea (1955), Callan et al. (1965) and Harrison (1965).

We recall the basics of Newtonian cosmology. We consider a spatially homogeneous solution of Eqs. (\ref{gpp4})-(\ref{gpp7}) of the form
\begin{equation}
\label{nc1}
\rho({\bf r},t)=\rho_b(t),\qquad {\bf u}({\bf r},t)=\frac{\dot a}{a}{\bf r},
\end{equation}
where $a(t)$ is the scale factor and $H=\dot a/a$ is the Hubble ``constant'' (in fact a function of time).
The Euler equations reduce to
\begin{equation}
\label{nc2}
\frac{d\rho_b}{dt}+3\rho_b\frac{\dot a}{a}=0,
\end{equation}
\begin{equation}
\label{nc3}
\frac{\ddot a}{a}{\bf r}=-\nabla\Phi_b.
\end{equation}
The pressure $p$ and the quantum potential $Q$ do not appear in the theory of the homogeneous model since they enter Eq. (\ref{gpp5}) only through their gradients. Equation (\ref{nc2}) leads to the relation
\begin{equation}
\label{nc4}
\rho_b a^3\sim 1,
\end{equation}
which corresponds to the conservation of mass. Taking the divergence of Eq. (\ref{nc3}) and using the Poisson equation (\ref{gpp6}), we obtain the cosmological equation
\begin{equation}
\label{nc5}
\frac{d^2 a}{dt^2}=-\frac{4}{3}\pi G \rho_b a+\frac{\Lambda}{3}a.
\end{equation}
Using Eq. (\ref{nc4}), its first integral is
\begin{equation}
\label{nc6}
\left (\frac{da}{dt}\right )^2=\frac{1}{3}(8\pi G\rho_b+\Lambda) a^2-\kappa,
\end{equation}
where $\kappa$ is a constant of integration. Equations (\ref{nc4})-(\ref{nc6}) are the Newtonian equations of an isotropic and homogeneous universe. They coincide with the equations derived by Friedmann (1922,1924) for $\Lambda=0$ and $\kappa=\pm 1$ and by Einstein \& de Sitter (1932) for $\Lambda=0$ and $\kappa=0$ from the theory of general relativity when the pressure is small compared to the energy density $\rho c^2$. In that case, $\kappa$ is the curvature constant and space is flat ($\kappa=0$), elliptical ($\kappa=1$) or hyperbolic ($\kappa=-1$)\footnote{Note that the constant $\kappa$ in Eq. (\ref{nc6}) can be set to unity by a suitable normalization of the parameters.}.

The Einstein (1917) static universe corresponds to
\begin{equation}
\label{nc7}
a=1, \qquad \rho_b=\frac{\Lambda}{4\pi G},\qquad \kappa=\Lambda.
\end{equation}
However, this universe is unstable against perturbations in $a$ (Eddington 1930, Harrison 1967). The Einstein-de Sitter (EdS) universe corresponds to $\Lambda=0$ and $\kappa=0$. In that case, Eqs. (\ref{nc5}) and (\ref{nc6}) reduce to
\begin{equation}
\label{nc8}
\ddot a=-\frac{4}{3}\pi G \rho_b a,\qquad {\dot a}^2=\frac{8}{3}\pi G\rho_b a^2.
\end{equation}
This yields
\begin{equation}
\label{nc9}
a\propto t^{2/3},\qquad H=\frac{\dot a}{a}=\frac{2}{3t}, \qquad \rho_b=\frac{1}{6\pi Gt^2}.
\end{equation}
Note that both inflationary theory (Guth 1981) and observations favor a flat universe ($\kappa=0$).
In that case, Eq. (\ref{nc6}) can be written $H^2=\frac{8}{3}\pi G(\rho_b+\rho_{\Lambda})$, where $\rho_{\Lambda}=\Lambda/8\pi G$ is the dark energy density, and it gives a relationship between Hubble's constant and the total density of the universe. With the present-day value of the Hubble
constant $H_0=2.273\, 10^{-18}\, {\rm s}^{-1}$, the Einstein-de Sitter
model ($\kappa=\Lambda=0$, $p\ll \rho c^2$) leads to an age of the universe $t_0=2/3H_0\sim
9.3$ billion years while the $\Lambda$CDM model, which is the standard
model of cosmology, predicts $13.75$ billion years. The effect of dark energy will be
considered in Secs. \ref{sec_contribution} and \ref{sec_dark}.

{\it Remark:} We can obtain equations (\ref{nc4})-(\ref{nc6}) directly if we use the ``naive" picture that the universe is a uniform sphere of radius $a(t)$, density $\rho_b(t)$ and mass $M$. The conservation of mass $M=\frac{4}{3}\pi \rho_b a^3$ leads to Eq. (\ref{nc4}). On the other hand, applying the Gauss theorem at the border of the sphere, Newton's equation can be written  ${d^2 a}/{dt^2}=-{GM}/{a^2}+{\Lambda}a/{3}$ leading to Eq. (\ref{nc5}). This is the equation of motion of a fictive particle of position $a$ in a potential $V(a)=-GM/a-\Lambda a^2/6$.  Its first integral is given by Eq. (\ref{nc6}) where $E=-\kappa/2$ can be regarded as the energy of the fictive particle. The case $E=0$ (EdS universe) corresponds to the situation where the velocity of the fictive particle is equal to the escape velocity. In this naive picture, $a$ may be interpreted as the ``radius'' of the universe. In fact, a better derivation (which does not assume a finite sphere) is to apply Newton's equation to an arbitrary fluid particle located in ${\bf r}$ and write ${d^2 r}/{dt^2}=-{G (4\pi\rho_b r^3/3)}/{r^2}+{\Lambda}r/{3}$. Setting ${\bf r}=a(t){\bf x}$ and dividing by $x$ yields Eq. (\ref{nc5}).

\section{Quantum Euler-Poisson system in an expanding universe}
\label{sec_qep}

We shall now study the instability of the homogeneous background and the growth of perturbations that ultimately lead to the large-scale structures of the universe. We shall work in the comoving frame (Peebles 1980). To that purpose, we set
\begin{equation}
\label{qep1}
{\bf r}=a(t){\bf x}, \qquad {\bf u}=\frac{\dot a}{a}{\bf r}+{\bf v},
\end{equation}
where ${\bf v}$ is the peculiar velocity. For the moment, we allow arbitrary deviations from the background flow. Let us first write the Poisson equation (\ref{gpp6}) in the comoving frame. Integrating Eq. (\ref{nc3}) and using Eq. (\ref{nc5}), we find that the background gravitational potential is
\begin{equation}
\label{qep2}
\Phi_b=-\frac{1}{2}\frac{\ddot a}{a}  r^2=-\frac{1}{2}{\ddot a} a x^2=\frac{2}{3}\pi G\rho_b(t)r^2-\frac{\Lambda}{6}r^2.
\end{equation}
This result can also be obtained by integrating the Poisson equation for a homogeneous system or by using the Gauss theorem. If we introduce the new potential $\phi=\Phi-\Phi_b$, i.e.
\begin{equation}
\label{qep3}
\phi=\Phi+\frac{1}{2}{\ddot a} a x^2,
\end{equation}
the Poisson equation (\ref{gpp6}) becomes
\begin{equation}
\label{qep4}
\Delta\phi=4\pi G a^2 (\rho-\rho_b),
\end{equation}
where the Laplacian is taken with respect to ${\bf x}$. Now, following Peebles (1980) and taking  the quantum pressure into account, we find that the hydrodynamic equations (\ref{gpp4}) and (\ref{gpp5}) can be written in the comoving frame as
\begin{equation}
\label{qep5}
\frac{\partial\rho}{\partial t}+\frac{3\dot a}{a}\rho+\frac{1}{a}\nabla\cdot (\rho {\bf v})=0,
\end{equation}
\begin{eqnarray}
\label{qep6}
\frac{\partial {\bf v}}{\partial t}+\frac{1}{a}({\bf v}\cdot \nabla){\bf v}+\frac{\dot a}{a}{\bf v}=-\frac{1}{\rho a}\nabla p-\frac{1}{a}\nabla\phi\nonumber\\
+\frac{\hbar^2}{2m^2a^3}\nabla \left (\frac{\Delta \sqrt{\rho}}{\sqrt{\rho}}\right ).
\end{eqnarray}
It is convenient to write the density in the form
\begin{eqnarray}
\label{qep7}
\rho=\rho_b(t)\left\lbrack 1+\delta({\bf x},t)\right \rbrack,
\end{eqnarray}
where $\rho_b\propto 1/a^3$ and $\delta({\bf x},t)$ is the density contrast (Peebles 1980). Substituting Eq. (\ref{qep7}) into Eqs.  (\ref{qep4})-(\ref{qep6}), we obtain the quantum barotropic Euler-Poisson system in an expanding universe
\begin{equation}
\label{qep8}
\frac{\partial\delta}{\partial t}+\frac{1}{a}\nabla\cdot ((1+\delta) {\bf v})=0,
\end{equation}
\begin{eqnarray}
\label{qep9}
\frac{\partial {\bf v}}{\partial t}+\frac{1}{a}({\bf v}\cdot \nabla){\bf v}+\frac{\dot a}{a}{\bf v}=-\frac{1}{\rho a}\nabla p-\frac{1}{a}\nabla\phi\nonumber\\
+\frac{\hbar^2}{2m^2a^3}\nabla \left (\frac{\Delta \sqrt{1+\delta}}{\sqrt{1+\delta}}\right ),
\end{eqnarray}
\begin{equation}
\label{qep10}
\Delta\phi=4\pi G \rho_b a^2 \delta.
\end{equation}
For $\hbar=p=0$, we recover the usual Euler equations of a cold gas ($T=0$) in an expanding universe (Peebles 1980)\footnote{If the universe is a classical collisionless fluid, the evolution of this fluid is fundamentally described by the Vlasov equation (Gilbert 1966). In that case, the hydrodynamic equations without pressure ($\hbar=p=0$) are based on some approximations (Peebles 1980) and they are not valid for all times. Alternatively, if dark matter is a BEC, the hydrodynamic equations (\ref{qep8})-(\ref{qep10}) are rigorously equivalent to the GPP system for all times.}. For a barotropic equation of state $p=p(\rho)$, the Euler equation (\ref{qep9}) can be rewritten
\begin{eqnarray}
\label{qep11}
\frac{\partial {\bf v}}{\partial t}+\frac{1}{a}({\bf v}\cdot \nabla){\bf v}+\frac{\dot a}{a}{\bf v}=-\frac{c_s^2}{(1+\delta)a}\nabla \delta-\frac{1}{a}\nabla\phi\nonumber\\
+\frac{\hbar^2}{2m^2a^3}\nabla \left (\frac{\Delta \sqrt{1+\delta}}{\sqrt{1+\delta}}\right ),
\end{eqnarray}
where $c_s^2=p'(\rho)=p'(\rho_b(1+\delta))$ is the square of the velocity of sound in the evolving system. It generically depends on position and time. For an isothermal equation of state $c_s^2=k_BT/m$ is a constant and for a polytropic equation of state $c_s^2=K\gamma\rho^{\gamma-1}$. For a standard BEC with a quartic self-interaction, described by the equation of state (\ref{gpp8}),  the square of the velocity of sound is
\begin{eqnarray}
\label{qep11b}
c_s^2=\frac{4\pi a_s\hbar^2\rho}{m^3},
\end{eqnarray}
and we obtain the system of equations
\begin{equation}
\label{qep12}
\frac{\partial\delta}{\partial t}+\frac{1}{a}\nabla\cdot ((1+\delta) {\bf v})=0,
\end{equation}
\begin{eqnarray}
\label{qep13}
\frac{\partial {\bf v}}{\partial t}+\frac{1}{a}({\bf v}\cdot \nabla){\bf v}+\frac{\dot a}{a}{\bf v}=-\frac{4\pi a_s\hbar^2\rho_b}{m^3 a}\nabla \delta\nonumber\\
-\frac{1}{a}\nabla\phi+\frac{\hbar^2}{2m^2a^3}\nabla \left (\frac{\Delta \sqrt{1+\delta}}{\sqrt{1+\delta}}\right ),
\end{eqnarray}
\begin{equation}
\label{qep14}
\Delta\phi=4\pi G \rho_b  a^2\delta.
\end{equation}

\section{Linearized equations}
\label{sec_lin}

Of course, $\delta=\phi=0$ and ${\bf v}={\bf 0}$ is a solution of the quantum Euler-Poisson system  (\ref{qep8})-(\ref{qep10}) in the comoving frame, corresponding to the pure background flow (\ref{nc1}). However, this solution may be unstable to some perturbations. If we consider small perturbations $\delta\ll 1$, $\phi\ll 1$, $|{\bf v}|\ll 1$ and linearize the foregoing equations we obtain
\begin{equation}
\label{lin1}
\frac{\partial\delta}{\partial t}+\frac{1}{a}\nabla\cdot {\bf v}=0,
\end{equation}
\begin{eqnarray}
\label{lin2}
\frac{\partial {\bf v}}{\partial t}+\frac{\dot a}{a}{\bf v}=-\frac{1}{a}c_s^2\nabla \delta-\frac{1}{a}\nabla\phi
+\frac{\hbar^2}{4m^2a^3}\nabla (\Delta\delta),
\end{eqnarray}
\begin{equation}
\label{lin3}
\Delta\phi=4\pi G \rho_b \delta a^2,
\end{equation}
where $c_s^2=p'(\rho_b(t))$ now denotes the square of the velocity of sound in the homogeneous background flow. It is just a function of time.  Taking the time derivative of Eq. (\ref{lin1}) multiplied by $a$, the divergence of Eq. (\ref{lin2}), and using Eq. (\ref{lin3}), these equations can be combined into a single equation governing the evolution of the density contrast
\begin{eqnarray}
\label{lin4}
\frac{\partial^2\delta}{\partial t^2}+2\frac{\dot a}{a}\frac{\partial\delta}{\partial t}=\frac{c_s^2}{a^2}\Delta\delta+4\pi G\rho_b \delta-\frac{\hbar^2}{4m^2a^4}\Delta^2\delta.
\end{eqnarray}
For $\hbar=0$, we recover the equation first derived by Bonnor (1957). In his case, the pressure $p$ is a kinetic pressure. Expanding the solution in Fourier modes of the form $\delta({\bf x},t)=\delta_{\bf k}(t)e^{i{\bf k}\cdot {\bf x}}$, we obtain
\begin{eqnarray}
\label{lin5}
\ddot\delta+2\frac{\dot a}{a}\dot\delta+\left (\frac{\hbar^2k^4}{4m^2a^4}+\frac{c_s^2k^2}{a^2}-4\pi G\rho_b\right )\delta=0,
\end{eqnarray}
where, for brevity,  we have noted $\delta(t)$ for $\delta_{\bf k}(t)$. For a standard BEC, using $c_s^2={4\pi a_s \hbar^2\rho_b}/{m^3}$, the foregoing equation can be rewritten
\begin{eqnarray}
\label{lin6}
\ddot\delta+2\frac{\dot a}{a}\dot\delta+\left (\frac{\hbar^2k^4}{4m^2a^4}+\frac{4\pi a_s \hbar^2 \rho_b k^2}{m^3a^2}-4\pi G\rho_b\right )\delta=0.\nonumber\\
\end{eqnarray}

\section{Quantum Jeans length}
\label{sec_qjl}

\subsection{Expanding universe}
\label{sec_e}

In the non-interacting case $a_s=c_s=0$, the equation for the density contrast reduces to
\begin{eqnarray}
\label{e1}
\ddot\delta+2\frac{\dot a}{a}\dot\delta+\left (\frac{\hbar^2k^4}{4m^2a^4}-4\pi G\rho_b\right )\delta=0.
\end{eqnarray}
From this relation, we can define a time-dependent quantum Jeans wavenumber
\begin{eqnarray}
\label{e2}
k_Q=\left (\frac{16\pi G\rho_b m^2a^4}{\hbar^2}\right )^{1/4}.
\end{eqnarray}
Note that the proper quantum Jeans wavenumber obtained by writing $\delta\propto e^{i{\bf k}_*\cdot {\bf r}}$ is $k_Q^*=({16\pi G\rho_b m^2}/{\hbar^2})^{1/4}$.
Recalling Eq. (\ref{nc4}), we can write $k_Q=\kappa_Q a^{1/4}$ where $\kappa_Q=({16\pi G\rho_b a^3 m^2}/{\hbar^2})^{1/4}$ is a constant. The quantum Jeans length $\lambda_Q=2\pi /k_Q$ decreases with time like $a^{-1/4}$ so that, in the comoving frame,  the system becomes unstable at smaller and smaller scales as the universe expands (note that, on the contrary, the proper quantum Jeans scale increases with time like $a^{3/4}$). In the cold, non-quantum universe, there is no Jeans length: all the scales are unstable. Therefore, quantum effects can stabilize the system at small scales and avoid the density cusps (Hu et al. 2000).

In the TF limit where the quantum potential can be neglected, the equation for the density contrast reduces to
\begin{eqnarray}
\label{e3}
\ddot\delta+2\frac{\dot a}{a}\dot\delta+\left (\frac{c_s^2k^2}{a^2}-4\pi G\rho_b\right )\delta=0.
\end{eqnarray}
From this relation, we can define a time-dependent classical Jeans wavenumber
\begin{eqnarray}
\label{e4}
k_J=\left (\frac{4\pi G\rho_b a^2}{c_s^2}\right )^{1/2}.
\end{eqnarray}
The proper Jeans wavenumber is $k_J^*=({4\pi G\rho_b}/{c_s^2})^{1/2}$. The evolution of $k_J(t)$ depends on the equation of state. For an isothermal equation of state for which $c_s^2=k_BT/m$, recalling Eq. (\ref{nc4}), we can write $k_J=\kappa_J a^{-1/2}$ where $\kappa_J=(4\pi G\rho_b a^3 m/k_BT)^{1/2}$ is a constant. The classical Jeans length $\lambda_J=2\pi/k_J$ increases with time like $a^{1/2}$ so that, in the comoving frame, the system becomes unstable at larger and larger scales as the universe expands. For a polytropic equation of state for which $c_s^2=K\gamma \rho_b^{\gamma-1}$, we can write $k_J=\kappa_J a^{(3\gamma-4)/2}$ where $\kappa_J=\lbrack 4\pi G(\rho_b a^3)^{2-\gamma}/K\gamma\rbrack^{1/2}$ is a constant. The classical Jeans wavelength behaves like $\lambda_J\propto a^{(4-3\gamma)/2}$. For $\gamma<4/3$, the Jeans wavelength grows with time and for $\gamma>4/3$, it decreases with time. For $\gamma=4/3$, the Jeans wavenumber is constant in time, i.e. $k_J=\kappa_J$. Finally, for a standard BEC, $c_s^2={4\pi a_s \hbar^2\rho_b}/{m^3}$, and the Jeans wavenumber can be written
\begin{eqnarray}
\label{e5}
k_J=\left (\frac{Gm^3a^2}{a_s\hbar^2}\right )^{1/2}.
\end{eqnarray}
It is independent on the density and can be written $k_J=\kappa_J a$ with $\kappa_J=(Gm^3/a_s\hbar^2)^{1/2}$. The Jeans length decreases like $\lambda_J\propto 1/a$.

{\it Remark:} the study of the Jeans instability in an expanding universe exhibits a critical value of the polytropic index $\gamma_{crit}=4/3$ (i.e. $n_{crit}=3$). It is interesting to note that the same critical index arises when one studies the dynamical stability of spatially inhomogeneous polytropic spheres with respect to the barotropic Euler-Poisson system. It has been established that a polytropic star is stable for $\gamma\ge 4/3$ and unstable otherwise (Binney \& Tremaine, 1987). Surprisingly, the same index appears in the stability analysis of a spatially homogeneous polytropic gas in an expanding universe. Note that the static study of Jeans (see the following section) does not directly exhibit a critical value of the polytropic index\footnote{The critical value $\gamma_{crit}=4/3$ only appears when one introduces the Jeans mass $M_J\sim \rho \lambda_J^3\propto \rho^{(3\gamma-4)/2}$, but it does not play any particular role in the stability analysis of the homogeneous fluid.}.

\subsection{Static universe}
\label{sec_s}

If we assume that Eq. (\ref{lin5}) remains valid in a static universe, and take $a=1$, we get
\begin{eqnarray}
\label{s1}
\ddot\delta+\left (\frac{\hbar^2k^4}{4m^2}+{c_s^2k^2}-4\pi G\rho_b\right )\delta=0.
\end{eqnarray}
Writing the perturbation in the form $\delta(t)\propto e^{-i\omega t}$, we obtain the dispersion relation
\begin{eqnarray}
\label{s2}
\omega^2=\frac{\hbar^2k^4}{4m^2}+{c_s^2k^2}-4\pi G\rho_b.
\end{eqnarray}
This dispersion relation has been studied in our previous paper (Chavanis 2011). The generalized Jeans wavenumber is given by
\begin{equation}
\label{s3}
k_c^2=\frac{2m^2}{\hbar^2}\left\lbrack \sqrt{c_s^4+\frac{4\pi G\hbar^2\rho_b}{m^2}}-c_s^2\right\rbrack.
\end{equation}
In the non-interacting case ($a_s=c_s=0)$, we recover the quantum Jeans wavenumber (\ref{e2}) with $a=1$ and in the TF limit, we recover the classical Jeans wavenumber (\ref{e4}) with $a=1$. The dispersion relation can be written $\omega^2/4\pi G\rho_b={k^4}/{k_Q^4}+{k^2}/{k_J^2}-1$ and the generalized Jeans wavenumber $k_c^2=({k_Q^4}/{2k_J^2})\lbrack \pm (1+{4k_J^4}/{k_Q^4})^{1/2}-1\rbrack$ with $+$ when $c_s^2\ge 0$ and $-$ when $c_s^2<0$. For $\lambda<\lambda_c$, $\omega$ is real and the perturbation oscillates with a pulsation $\omega$; for $\lambda>\lambda_c$, $\omega$ is purely imaginary and the perturbation grows exponentially rapidly with a growth rate $\gamma=\sqrt{-\omega^2}$. Jeans' stability criterion is therefore $\lambda<\lambda_c$. For $c_s^2>0$, the maximum growth rate corresponds to $k=0$ (infinite wavelengths) and is given by $\gamma=\sqrt{4\pi G\rho_b}$. On the other hand, considering a  BEC with negative scattering length, it is found (Chavanis 2011) that the maximum growth rate corresponds to $k_*=(8\pi |a_s|\rho_b/m)^{1/2}$ and is given by
\begin{eqnarray}
\label{s4}
\gamma_*=\sqrt{\frac{16\pi^2a_s^2\hbar^2\rho_b^2}{m^4}+4\pi G\rho_b}.
\end{eqnarray}
Note that $k_*=({k_Q^4}/{2|k_J^2|})^{1/2}$ and $\gamma_*=\sqrt{4\pi G\rho} (1+{k_Q^4}/{4k_J^4})^{1/2}$. For $|a_s|\gg (Gm^4/4\pi\hbar^2\rho_b)^{1/2}$, we find that $\gamma_*\sim 4\pi|a_s|\hbar\rho_b/m^2\gg  \sqrt{4\pi G\rho_b}$. Therefore, an attractive short-range interaction ($a_s<0$) increases the growth rate of the Jeans instability.

Jeans' classical analysis (Jeans 1902, 1929) suffers from the defect that in general there is no initial stationary state that is in a uniform non-rotating fluid. Therefore, using Eq. (\ref{s1}) in a static universe is referred to as the ``Jeans swindle''\footnote{Kiessling (2003) provides a vindication of the ``Jeans swindle''. He argues that, when considering an infinite and homogeneous distribution of matter, the Poisson equation must be modified so as to correctly define the gravitational force. He proposes to use a regularization of the form $\Delta\Phi-k_0^2\Phi=4\pi G\rho$ where $k_0$ is an inverse screening length that ultimately tends to zero ($k_0\rightarrow 0$), or a regularization of the form  $\Delta\Phi=4\pi G(\rho-\overline{\rho})$ where $\overline{\rho}$ is the mean density. In his point of view, this is not a swindle but just the right way to make the problem mathematically rigorous and have a well-defined gravitational force.}. As noted by Bonnor (1957), the Jeans procedure is only valid in the static Einstein universe\footnote{As is well-known, Einstein (1917) introduced a cosmological constant in the equations of general relativity in order to recover a homogeneous static universe. He also considered, as a preamble of his paper, a modification of the classical Poisson equation in the form $\Delta\Phi-k_0^2\Phi=4\pi G\rho$ because he (incorrectly) believed that, in the Newtonian world model, the cosmological constant is equivalent to a screening length (see Spiegel 1998, Kiessling 2003 and Chavanis \& Delfini 2010 for historical details).}. However, in this last case, the background density of the universe is $\rho_b=\Lambda/4\pi G$. The justification does not apply to a uniform mass of gas of different density. Furthermore, the Einstein universe is strongly unstable. Therefore, this justification is not valid and it is necessary to develop the Jeans instability analysis in an expanding universe.

{\it Remark:} Jeans' analysis could be valid if the growth rate of the instability were much larger than the rate of the expansion of the universe. According to Eq. (\ref{nc8}), the rate of expansion of the universe is $H=\dot a/a=(8\pi G\rho/3)^{1/2}$. On the other hand, when $c_s^2>0$, the maximum growth rate of the Jeans instability is $\gamma=\sqrt{4\pi G\rho}$. They are exactly of the same order which leads to the usual conclusion that Jeans' instability analysis is not valid (see, e.g., Weinberg 1972). However, when $c_s^2<0$, the maximum growth rate, given by Eq. (\ref{s4})  can be much larger. In that case, the expansion of the universe could be ignored and  the static Jeans instability analysis could be valid.

\section{Solution in the Einstein-de Sitter universe}
\label{sec_eds}

The evolution of the perturbations is more complicated to analyze in an expanding universe than in a static universe (and turns out to be very different). For simplicity, we consider the case of an Einstein-de Sitter universe. In that case, it is possible to obtain analytical solutions of the linearized equation (\ref{lin5}) for the density contrast  in some particular cases. Measuring the evolution in terms of $a$ instead of $t$ and using Eq. (\ref{nc8}),  Eq. (\ref{lin5}) is transformed into
\begin{eqnarray}
\label{eds1}
\frac{d^2\delta}{da^2}+\frac{3}{2a}\frac{d\delta}{da}+\frac{3}{2a^2}\left (\frac{\hbar^2k^4}{16\pi G\rho_b m^2a^4}+\frac{c_s^2k^2}{4\pi G\rho_b a^2}-1\right )\delta=0.\nonumber\\
\end{eqnarray}

For $\hbar=c_s=0$, we recover the case of a {\it cold classical universe} in which the pressure is zero and quantum effects are neglected. Equation (\ref{eds1}) reduces to
\begin{eqnarray}
\label{eds2}
\frac{d^2\delta}{da^2}+\frac{3}{2a}\frac{d\delta}{da}-\frac{3}{2a^2}\delta=0,
\end{eqnarray}
and its  solutions are
\begin{eqnarray}
\label{eds3}
\delta_+\propto a, \qquad \delta_-\propto a^{-3/2}.
\end{eqnarray}
We note that the growth of the perturbations is algebraic in an expanding universe while the ``naive'' Jeans instability analysis in a static universe predicts an exponential growth. Therefore, the results of the two analysis are very different. As mentioned long ago by Bonnor (1957) and others, this (slow) algebraic growth may be a problem to form large-scale structures sufficiently rapidly.

Assuming a polytropic equation of state $p=K\rho^{\gamma}$ for which $c_s^2=K\gamma\rho_b^{\gamma-1}$ and introducing the variables of Sec. \ref{sec_e}, we can rewrite Eq. (\ref{eds1}) in the form
\begin{eqnarray}
\label{eds4}
\frac{d^2\delta}{da^2}+\frac{3}{2a}\frac{d\delta}{da}+\frac{3}{2a^2}\left (\frac{k^4}{\kappa_Q^4 a}+\frac{k^2}{\kappa_J^2a^{3\gamma-4}}-1\right )\delta=0.\quad
\end{eqnarray}
Making the change of variables $\delta(a)=f(a)/a$, we find that the differential equation for $f$ is
\begin{eqnarray}
\label{eds5}
\frac{d^2 f}{da^2}-\frac{1}{2a}\frac{df}{da}+\frac{1}{a^2}\left (\frac{3k^4}{2\kappa_Q^4 a}+\frac{3k^2}{2\kappa_J^2 a^{3\gamma-4}}-1\right )f=0.\quad
\end{eqnarray}
We have not been able to find the general solution of this equation in
terms of simple functions so that we shall consider particular
cases. Some of the solutions presented below have already been
discussed in the literature (see, e.g., Harrison 1967) but we shall
give  more detail and put emphasis on the
solutions describing self-gravitating BECs that were not discussed
previously.

\subsection{The TF approximation}
\label{sec_tf}

In the TF approximation where the quantum potential can be neglected, Eq. (\ref{eds5}) reduces to
\begin{eqnarray}
\label{tf1}
\frac{d^2 f}{da^2}-\frac{1}{2a}\frac{df}{da}+\frac{1}{a^2}\left (\frac{3k^2}{2\kappa_J^2a^{3\gamma-4}}-1\right )f=0.
\end{eqnarray}
This is a particular case of  the equation studied by Savedoff \& Vila (1962) in relation to the work of Bonnor (1957)\footnote{In this section, we allow $p$ to have the interpretation of an ordinary pressure, i.e. we consider not only a gas of BECs but also a more conventional gas with a kinetic pressure.}. It can be solved in terms of Bessel functions (see Appendix \ref{sec_sol}). For $\gamma\neq 4/3$, $f(a)$ is given by Eq. (\ref{sol7}) and the evolution of the density contrast is
\begin{eqnarray}
\label{tf2}
\delta(a)\propto \frac{1}{a^{1/4}}J_{\pm \frac{5}{2(4-3\gamma)}}\left (\frac{\sqrt{6}}{4-3\gamma} \frac{k}{\kappa_J} a^{\frac{4-3\gamma}{2}}\right ).
\end{eqnarray}
Using the asymptotic expansions of the Bessel functions, we note that
the density contrast is oscillating for $k\gg k_J(a)$ and growing for
$k\ll k_J(a)$ (on the timescale over which these inequalities are fulfilled). 
Therefore, wavelengths which are long with respect to
$\lambda_J(a)$ behave like in a cold universe whereas for the short
wavelengths the perturbations have an oscillatory behavior. In this
sense, the results are similar to those obtained with the classical
Jeans analysis. They are in fact different because (i) the Jeans
length varies with time, (ii) the oscillating solutions can
be growing or decaying and (iii) the growth is algebraic instead
of exponential. In fact, the results more crucially depend on the
value of the polytropic index $\gamma$ than on the Jeans length
$\lambda_J$. We must distinguish four cases. (i) $\gamma<4/3$: for
$a\rightarrow 0$ the perturbations grow like in a cold gas
($\delta_+\sim a$) and, for $a\rightarrow +\infty$, they undergo damped
oscillations (they decay like $1/a^{(5-3\gamma)/4}$); (ii)
$4/3<\gamma<5/3$: for $a\rightarrow 0$ the perturbations diverge like
$1/a^{(5-3\gamma)/4}$ while oscillating implying that we must start
the stability analysis at $a_i>0$. For $a\rightarrow a_i$, the
perturbations undergo damped oscillations (they decay like
$1/a^{(5-3\gamma)/4}$) and, for $a\rightarrow +\infty$, they grow like
in a cold gas ($\delta_+\sim a$); (iii) $\gamma=5/3$: for
$a\rightarrow 0$ the perturbations oscillate and, for $a\rightarrow
+\infty$, they grow like in a cold gas ($\delta_+\sim a$); (iv)
$\gamma>5/3$: for $a\rightarrow 0$ the perturbations undergo growing
oscillations (they grow like $a^{(3\gamma-5)/4}$) and for
$a\rightarrow +\infty$ they grow like in a cold gas ($\delta_+\sim
a$).  Considering the regime $a\ll 1$, the perturbations grow
for $\gamma<4/3$, decay for $4/3<\gamma<5/3$ and grow for
$\gamma>5/3$. Considering the regime $a\gg 1$, the perturbations
decay for $\gamma<4/3$ and grow for $\gamma>4/3$. For
$\gamma<4/3$, the perturbations start to grow but finally decay. In a
sense, the system is asymptotically stable. However, the growth of the
perturbations in the initial stage can trigger nonlinear effects that
may induce instabilities.

For $\gamma=4/3$ ($n=3$), the density contrast behaves like
\begin{eqnarray}
\label{tf3}
\delta(a)\propto a^{-\frac{1}{4}\pm \sqrt{\frac{25}{16}-\frac{3k^2}{2\kappa_J^2}}}.
\end{eqnarray}
This polytropic index corresponds to a gas of photons at temperature $T$ or to a gas of relativistic  fermions at zero temperature. In that case, the Jeans length is independent on time: $k_J=\kappa_J$. For $k>(25/24)^{1/2}k_J$ the density contrast behaves like
\begin{eqnarray}
\label{tf4}
\delta(a)\propto \frac{1}{a^{1/4}}\cos\left (\sqrt{\frac{3k^2}{2\kappa_J^2}-\frac{25}{16}}\ln a+\phi\right ).
\end{eqnarray}
It diverges for $a\rightarrow 0$ implying that we must start the stability analysis at $a_i>0$. For $a\ge a_i$, the perturbations decay like ${a^{-1/4}}$ while making oscillations. For $k_J<k<(25/24)^{1/2}k_J$, the perturbations decay algebraically without oscillating. For $k<k_J$, the perturbations grow algebraically (more precisely $\delta_+$ grows and $\delta_-$ decays). For $k\rightarrow 0$, the perturbations behave like in a cold gas ($\delta_+$ grows like $a$ and $\delta_-$ decays like $a^{-2/3}$). This situation is relatively close to the classical Jeans stability analysis but (i) the growth of the perturbation is algebraic instead of exponential; (ii) the oscillatory solutions decay with time; (iii)  there is a small interval $k_J<k<(25/24)^{1/2}k_J$ that has no counterpart in the static Jeans analysis.

Let us consider particular  equations of state with a physical meaning. The index $\gamma=1$ corresponds to an {\it isothermal universe}. The density contrast is given by
\begin{eqnarray}
\label{tf5}
\delta(a)\propto \frac{1}{a^{1/4}}J_{\pm \frac{5}{2}}\left ({\sqrt{6}} \frac{k}{\kappa_J} a^{{1}/{2}}\right ).
\end{eqnarray}
For the index $\gamma=5/3$ ($n=3/2$), we obtain
\begin{eqnarray}
\label{tf6}
\delta(a)\propto \frac{1}{a^{1/4}}J_{\pm \frac{5}{2}}\left (\sqrt{6} \frac{k}{\kappa_J} \frac{1}{a^{{1}/{2}}}\right ).
\end{eqnarray}
This polytropic index corresponds to a gas of non-relativistic fermions at zero temperature.  For the polytropic index $\gamma=2$ ($n=1$), we have
\begin{eqnarray}
\label{tf7}
\delta(a)\propto \frac{1}{a^{1/4}}J_{\pm \frac{5}{4}}\left (\sqrt{\frac{3}{2}} \frac{k}{\kappa_J} \frac{1}{a}\right ).
\end{eqnarray}
This polytropic index corresponds to a  BEC with quartic self-interaction in the TF approximation (see Sec. \ref{sec_gpp}). In that case, $\kappa_J=(Gm^3/a_s\hbar^2)^{1/2}$. The previous expression assumes that $a_s>0$. In the case where $a_s<0$, we obtain
\begin{eqnarray}
\label{tf8}
\delta(a)\propto \frac{1}{a^{1/4}}I_{\pm \frac{5}{4}}\left (\sqrt{\frac{3}{2}}\frac{k}{\kappa_J} \frac{1}{a}\right ).
\end{eqnarray}
with $\kappa_J=(Gm^3/|a_s|\hbar^2)^{1/2}$. For $a\rightarrow +\infty$, the perturbations behave like in a cold gas. For $a\rightarrow 0$, we find that
\begin{eqnarray}
\label{tf9}
\delta(a)\propto a^{1/4}e^{\sqrt{\frac{3}{2}}\frac{k}{k_J}\frac{1}{a}}\rightarrow +\infty.
\end{eqnarray}
In a sense, this exponential divergence demonstrates that an attractive self-interaction accelerates the growth of the perturbations. In that case, we can neglect the expansion of the universe and we are led back to the static study of Sec. \ref{sec_s}.

\begin{figure}[!h]
\begin{center}
\includegraphics[clip,scale=0.3]{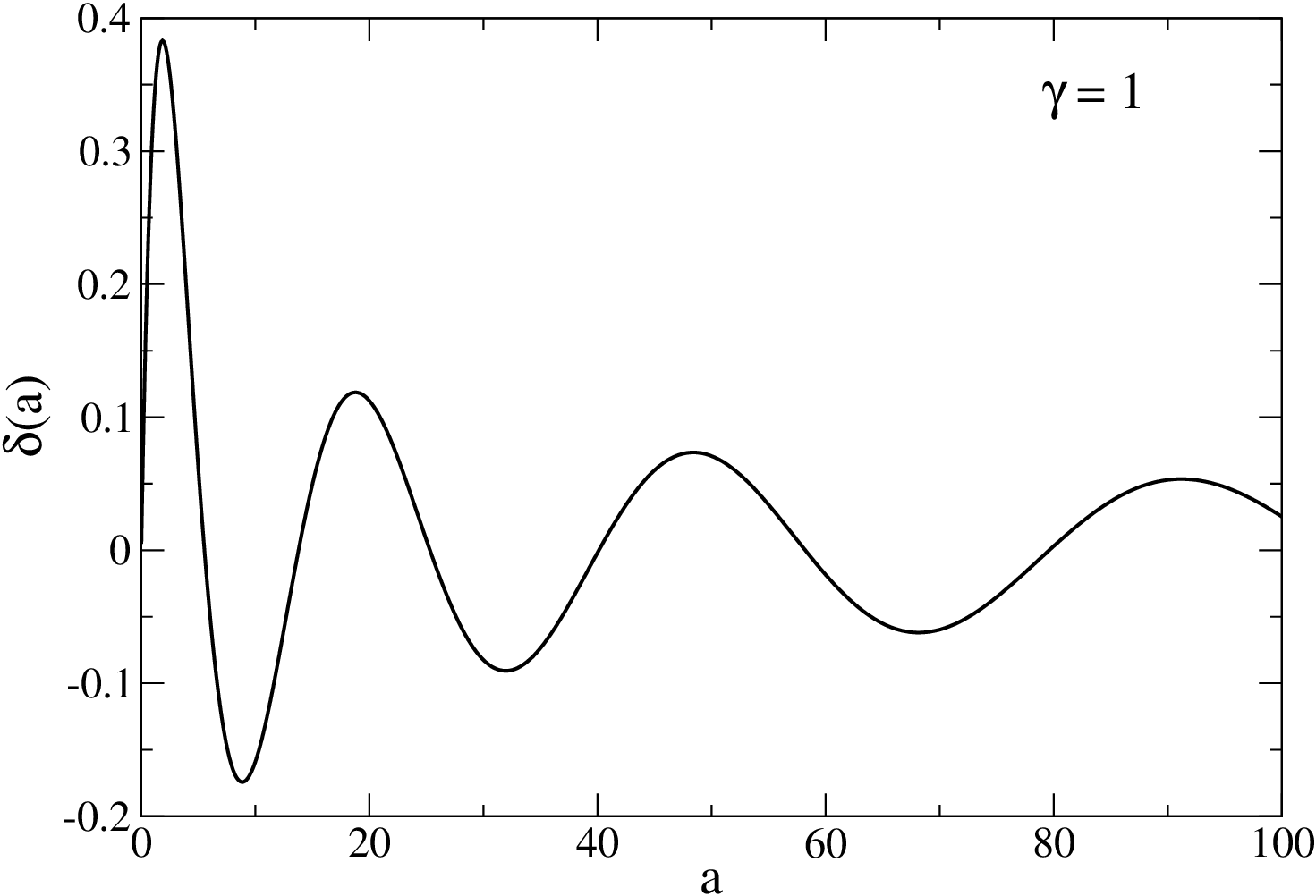}
\caption{Evolution of the perturbation $\delta(a)$ in an isothermal universe ($\gamma=1$). This corresponds to the case $\gamma<4/3$.  We have taken $k=\kappa_J$.}
\label{gamma1}
\end{center}
\end{figure}

\begin{figure}[!h]
\begin{center}
\includegraphics[clip,scale=0.3]{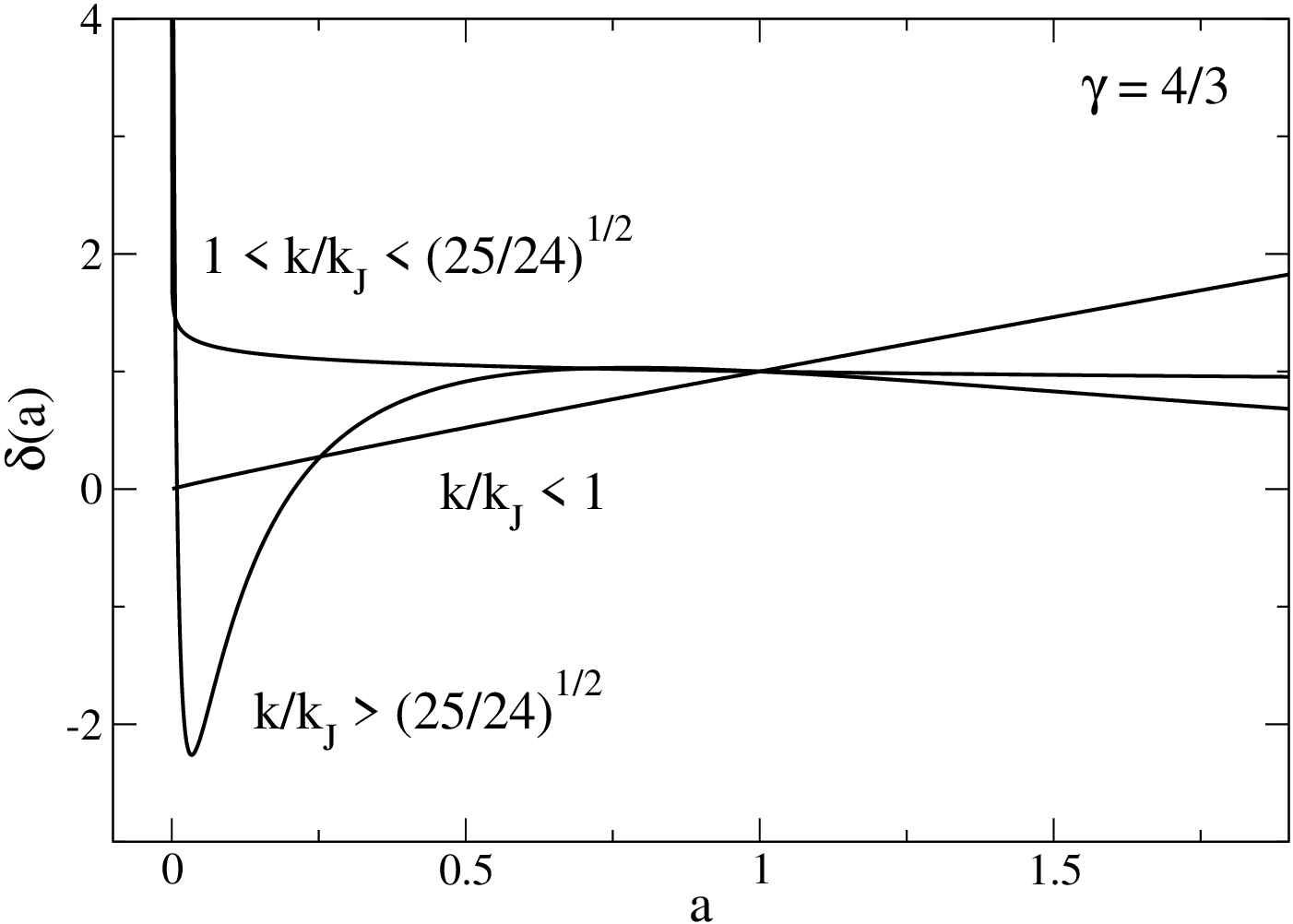}
\caption{Evolution of the perturbation $\delta(a)$ in a photonic or in a relativistic fermionic  universe ($\gamma=4/3$). We have represented the solution $\delta_+$.}
\label{gamma4sur3}
\end{center}
\end{figure}

\begin{figure}[!h]
\begin{center}
\includegraphics[clip,scale=0.3]{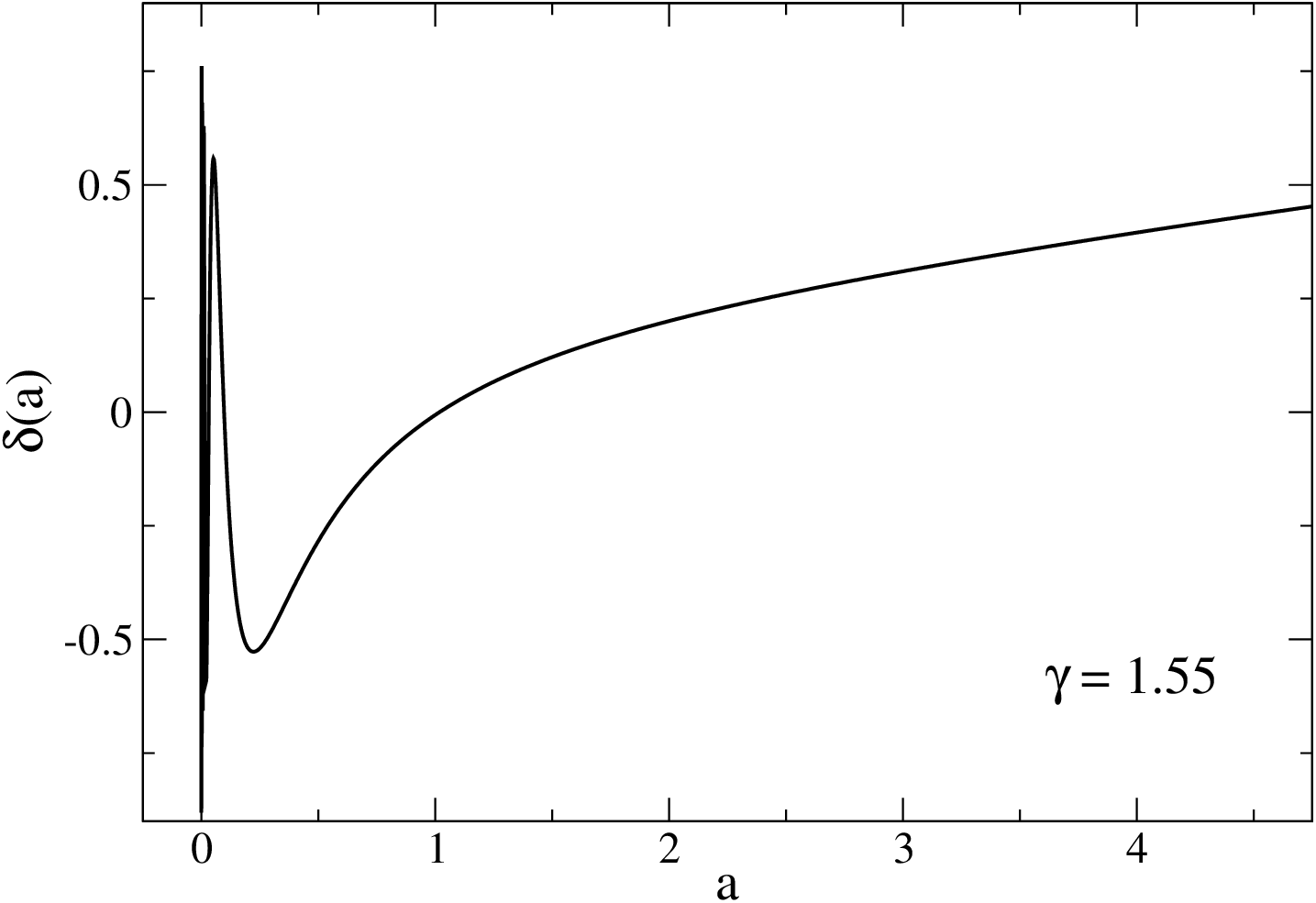}
\caption{Evolution of the perturbation $\delta(a)$ in a polytropic universe with $\gamma=1.55$. This corresponds to the case $4/3<\gamma<5/3$.  We have taken $k=\kappa_J$.}
\label{gamma1.55}
\end{center}
\end{figure}

\begin{figure}[!h]
\begin{center}
\includegraphics[clip,scale=0.3]{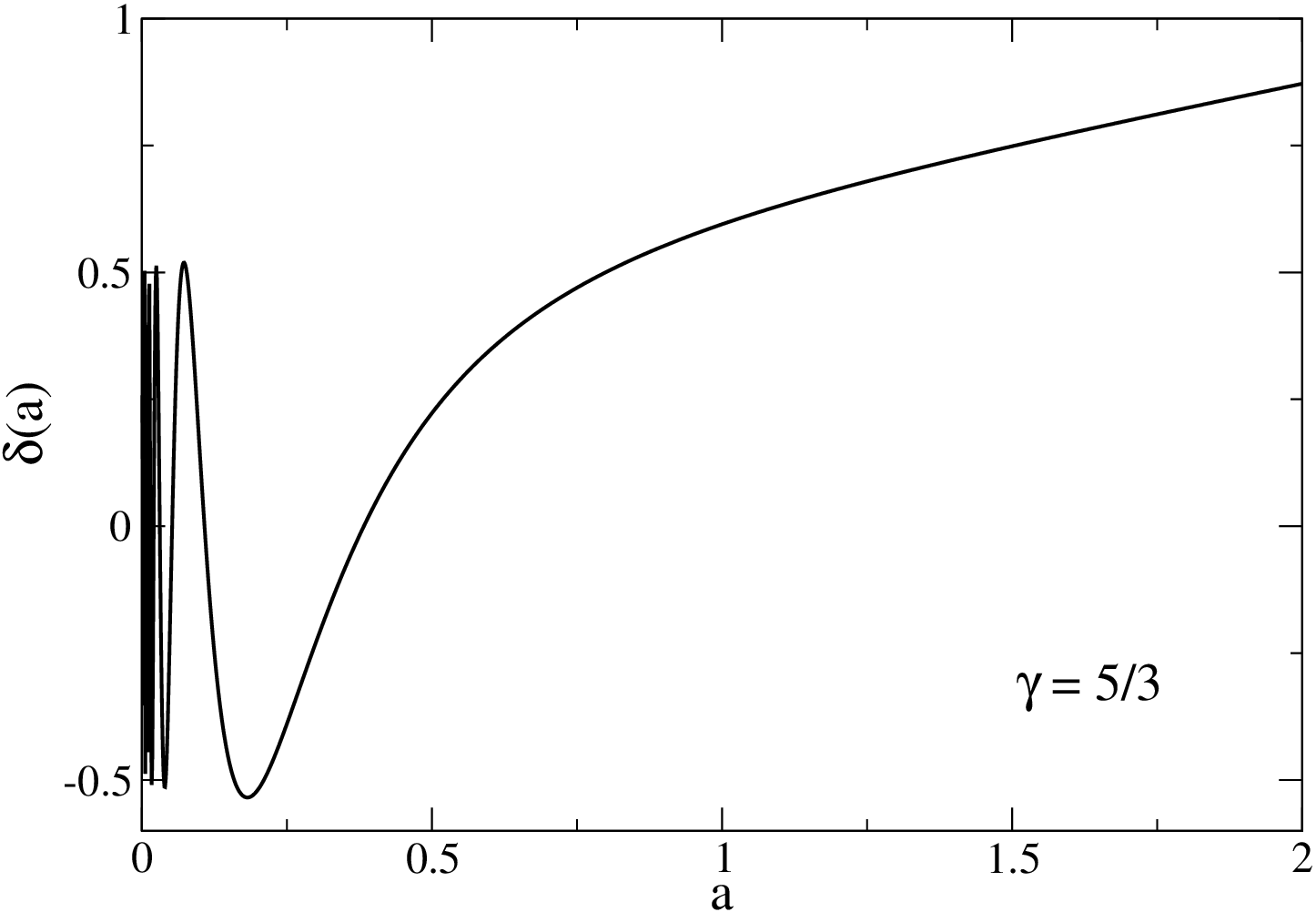}
\caption{Evolution of the perturbation $\delta(a)$ in a non-relativistic fermionic universe or in a BEC universe without self-interaction ($\gamma=5/3$). We have taken $k=\kappa_J$.}
\label{gamma5sur3}
\end{center}
\end{figure}

\begin{figure}[!h]
\begin{center}
\includegraphics[clip,scale=0.3]{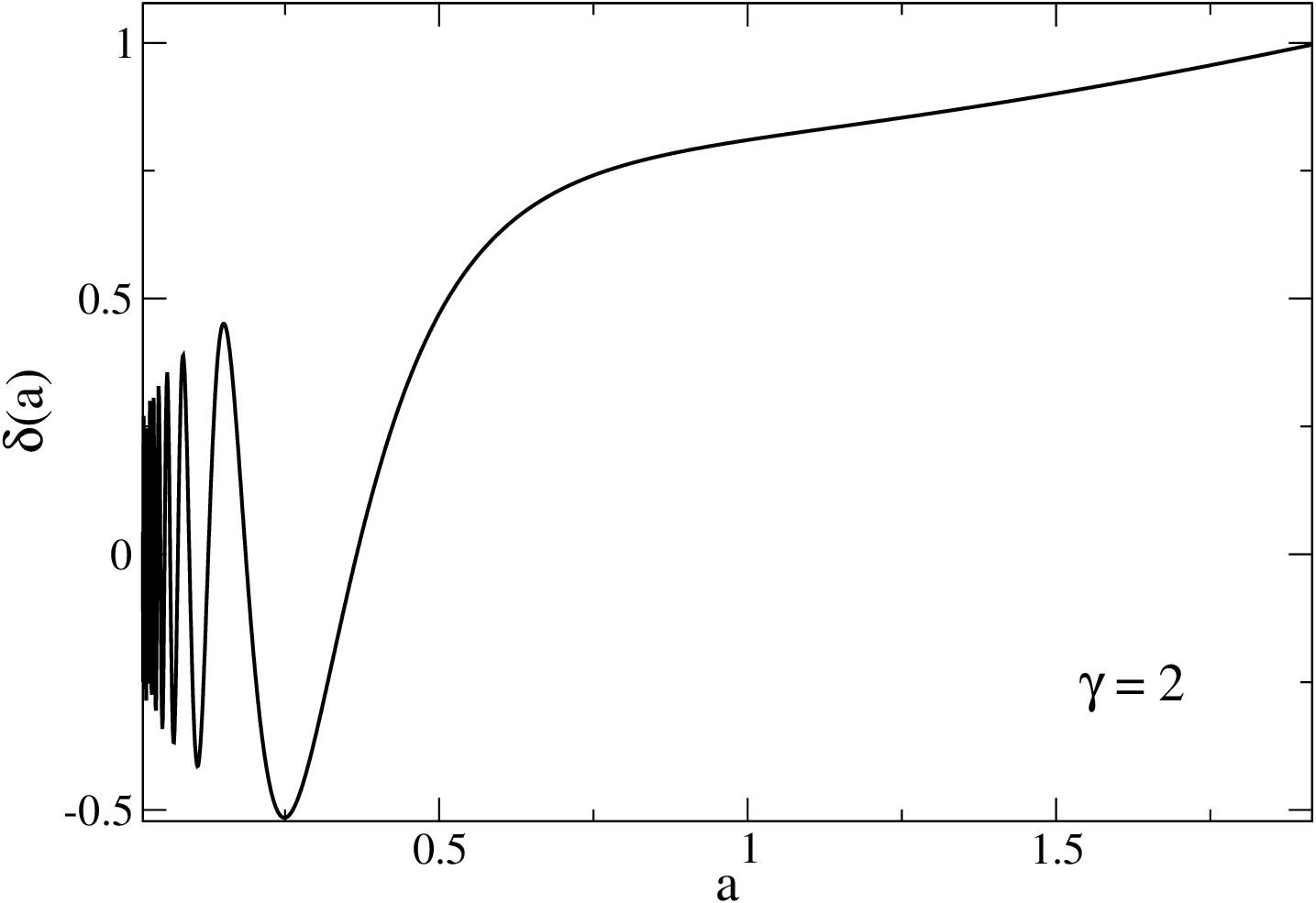}
\caption{Evolution of the perturbation $\delta(a)$ in a BEC universe with quartic self-interaction in the TF limit ($\gamma=2$).  This corresponds to the case $\gamma>5/3$. We have taken $k=\kappa_J$. }
\label{gamma2}
\end{center}
\end{figure}

Some curves representing the evolution of the density contrast $\delta(a)$ in the different cases listed above are represented in Figs. \ref{gamma1}-\ref{gamma2} for illustration.

\subsection{The non-interacting BEC case}
\label{sec_ni}

In the non-interacting case $a_s=c_s=0$, which corresponds to a {\it BEC universe} without self-interaction, Eq. (\ref{eds5}) reduces to
\begin{eqnarray}
\label{ni1}
\frac{d^2 f}{da^2}-\frac{1}{2a}\frac{df}{da}+\frac{1}{a^2}\left (\frac{3k^4}{2\kappa_Q^4 a}-1\right )f=0.
\end{eqnarray}
Comparing this equation with Eq. (\ref{tf1}), we see that the dependance in $a$ is the same as for a polytrope $\gamma=5/3$.  Therefore, $f(a)$ is given by Eq. (\ref{sol7}) with $\lambda={3k^4}/{2\kappa_Q^4}$ and $\alpha=-1/2$ yielding
\begin{eqnarray}
\label{ni2}
\delta(a)\propto \frac{1}{a^{1/4}}J_{\pm \frac{5}{2}}\left (\sqrt{6} \frac{k^2}{\kappa_Q^2} \frac{1}{a^{{1}/{2}}}\right ).
\end{eqnarray}
For $a\rightarrow 0$ the perturbations oscillate and for $a\rightarrow +\infty$ they behave like in a cold gas ($\delta_+$ grows like $a$). In a sense, a cold bosonic universe behaves similarly to a cold non-relativistic  fermionic universe (compare Eqs. (\ref{ni2}) and (\ref{tf6})). However, the dependence in $k$ is different ($k^2$ instead of $k$) and the expression of the Jeans length also is different (compare Eqs. (\ref{e2}) and (\ref{e4})).

\subsection{The non-relativistic fermionic case $\gamma=5/3$}
\label{sec_fermi}

A non-relativistic gas of fermions at $T=0$ is usually described by a polytropic equation of state $p=K\rho^{\gamma}$ with index $\gamma=5/3$ ($n=3/2$) and polytropic constant $K=(1/5)(3/8\pi)^{2/3}h^2/m^{8/3}$ (Chandrasekhar 1939). This equation of state arises from the Pauli exclusion principle. Such a semi-classical description is valid in the TF approximation which is exact when $N\rightarrow +\infty$. In some cases, it can be relevant to go beyond the TF approximation and take into account the quantum pressure arising from the Heisenberg principle. This can regularize the dynamics at small scales as shown by Bilic et al. (2001) in their study of the formation of fermion stars by gravitational collapse. In our problem, this general situation is described by Eq. (\ref{eds4}) with $\gamma=5/3$. This yields
\begin{eqnarray}
\label{fermi1}
\frac{d^2\delta}{da^2}+\frac{3}{2a}\frac{d\delta}{da}+\frac{3}{2a^2}\left (\frac{k^4}{\kappa_Q^4 a}+\frac{k^2}{\kappa_J^2a}-1\right )\delta=0.
\end{eqnarray}
The equation for $f(a)$ is
\begin{eqnarray}
\label{fermi2}
\frac{d^2 f}{da^2}-\frac{1}{2a}\frac{df}{da}+\frac{1}{a^2}\left (\frac{3k^4}{2\kappa_Q^4a}+\frac{3k^2}{2\kappa_J^2 a}-1\right )f=0.
\end{eqnarray}
As noted in Sec. \ref{sec_ni},  the two terms have the same dependence on $a$. Therefore, $f(a)$ is given by Eq. (\ref{sol7}) with
\begin{eqnarray}
\label{fermi3}
\lambda=\frac{3k^4}{2\kappa_Q^4}+\frac{3k^2}{2\kappa_J^2},\qquad \alpha=-\frac{1}{2}.
\end{eqnarray}
We obtain
\begin{eqnarray}
\label{fermi4}
\delta(a)\propto \frac{1}{a^{1/4}}J_{\pm \frac{5}{2}}\left (\sqrt{\frac{6k^4}{\kappa_Q^4}+\frac{6k^2}{\kappa_J^2}} \frac{1}{a^{{1}/{2}}}\right ).
\end{eqnarray}
For $a\rightarrow 0$ the perturbations oscillate and, for $a\rightarrow +\infty$, they behave like in a cold gas.

\subsection{The relativistic fermionic case $\gamma=4/3$}
\label{sec_fermirel}

A relativistic gas of fermions at $T=0$ is usually described by a polytropic equation of state $p=K\rho^{\gamma}$ with index $\gamma=4/3$ ($n=3$) and polytropic constant $K=(1/4)(3/8\pi)^{1/3}hc/m^{4/3}$ (Chandrasekhar 1939). If we go beyond the TF approximation and take  the quantum pressure into account, the evolution of the density contrast is described by Eq. (\ref{eds4}) with $\gamma=4/3$. This yields
\begin{eqnarray}
\label{frel1}
\frac{d^2\delta}{da^2}+\frac{3}{2a}\frac{d\delta}{da}+\frac{3}{2a^2}\left (\frac{k^4}{\kappa_Q^4 a}+\frac{k^2}{\kappa_J^2}-1\right )\delta=0.\quad
\end{eqnarray}
The equation for $f(a)$ is
\begin{eqnarray}
\label{frel2}
\frac{d^2 f}{da^2}-\frac{1}{2a}\frac{df}{da}+\frac{1}{a^2}\left (\frac{3k^4}{2\kappa_Q^4 a}+\frac{3k^2}{2\kappa_J^2}-1\right )f=0.\quad
\end{eqnarray}
Its solution is given by Eq. (\ref{sol13}) and the evolution of the density contrast is
\begin{eqnarray}
\label{frel3}
\delta(a)\propto \frac{1}{a^{1/4}}J_{\pm \frac{1}{2}\sqrt{25-24\frac{k^2}{\kappa_J^2}}}\left (\sqrt{6} \frac{k^2}{\kappa_Q^2} \frac{1}{a^{1/2}}\right ).
\end{eqnarray}
We note that the classical Jeans scale appears in the index of the Bessel function while the quantum Jeans scale appears in the argument of the Bessel function.

\subsection{The general case (asymptotics)}
\label{sec_genasy}

We now treat the general case, taking into account quantum pressure
and polytropic pressure (we assume $a_s>0$). Since we cannot solve
Eq. (\ref{eds4}) analytically, we just give asymptotic results.

Let us first consider the limit $a\rightarrow 0$. (i) If $\gamma<5/3$, the polytropic pressure is negligible in front of the quantum pressure and the perturbations oscillate (see Sec. \ref{sec_ni}). (ii)  If  $\gamma=5/3$, the polytropic pressure and  the quantum pressure are of the same order, and the perturbations oscillate (see Sec. \ref{sec_fermi}). (iii)  If $\gamma>5/3$, the quantum pressure is negligible in front of the polytropic pressure and the perturbations undergo growing oscillations  $\delta\propto a^{(3\gamma-5)/4}$ (see Sec. \ref{sec_tf}). Therefore, the perturbations oscillate for $\gamma\le 5/3$ and grow for $\gamma>5/3$. The quantum pressure has a stabilizing role for $\gamma\le 4/3$ (see Sec. \ref{sec_tf}).

Let us now consider the limit $a\rightarrow +\infty$. (i) If $\gamma<4/3$, the quantum pressure is negligible in front of the polytropic pressure and the perturbations undergo damped oscillations and decay like $1/a^{(5-3\gamma)/4}$ (see Sec. \ref{sec_tf}). (ii) If $\gamma=4/3$, the quantum pressure is negligible in front of the polytropic pressure and we are led back to the critical case of Sec. \ref{sec_tf}: for $k>(25/24)^{1/2}k_J$ the perturbations  decay like ${a^{-1/4}}$ while making oscillations, for $k_J<k<(25/24)^{1/2}k_J$ the perturbations decay algebraically without oscillating and  for $k<k_J$ the perturbations grow algebraically. (iii) For $\gamma>4/3$, the perturbations behave like in a cold gas and grows like $\delta_+\sim a\rightarrow +\infty$. Therefore, the perturbations decay for $\gamma<4/3$ and grow for $\gamma>4/3$.

\section{Bose-Einstein condensate universe}
\label{sec_pressure}

\subsection{Newtonian cosmology with pressure}
\label{sec_basic}

We now take into account pressure effects (coming from special relativity) in the evolution of the cosmic fluid. We base our study on the set of hydrodynamic equations
\begin{equation}
\label{b1}
\frac{\partial\rho}{\partial t}+\nabla\cdot (\rho {\bf u})+\frac{p}{c^2}\nabla\cdot {\bf u}=0,
\end{equation}
\begin{equation}
\label{b2}
\frac{\partial {\bf u}}{\partial t}+({\bf u}\cdot \nabla){\bf u}=-\frac{\nabla p}{\rho+\frac{p}{c^2}}-\nabla\Phi,
\end{equation}
\begin{equation}
\label{b3}
\Delta\Phi=4\pi G \left (\rho+\frac{3p}{c^2}\right )-\Lambda,
\end{equation}
where $c$ is the velocity of light. For $c\rightarrow +\infty$, we recover the classical Euler-Poisson system (Binney \& Tremaine 1987). These equations were introduced  by McCrea (1951). His initial continuity equation, which contained an incorrect pressure gradient term, was corrected by Lima {et al.} (1997) and we took this correction into account in writing Eq. (\ref{b1}). These equations lead to the correct relativistic equations for the cosmic evolution. Indeed, assuming a homogeneous and isotropic solution of the form $\rho({\bf r},t)=\rho_b(t)$, $p({\bf r},t)=p_b(t)$ and ${\bf u}({\bf r},t)=(\dot a/a){\bf r}$, Eqs. (\ref{b1})-(\ref{b3}) reduce to
\begin{equation}
\label{b5}
\frac{d\rho_b}{dt}+3\frac{\dot a}{a}\left (\rho_b+\frac{p_b}{c^2}\right )=0,
\end{equation}
\begin{equation}
\label{b6}
\frac{\ddot a}{a}=-\frac{4\pi G}{3} \left (\rho_b+\frac{3p_b}{c^2}\right )+\frac{\Lambda}{3},
\end{equation}
\begin{equation}
\label{b7}
\left (\frac{\dot a}{a}\right )^2=\frac{8\pi G}{3}\rho_b-\frac{\kappa}{a^2}+\frac{\Lambda}{3}.
\end{equation}
These are precisely the Friedmann equations that can be derived from the theory of general relativity (Weinberg 1972). The perturbation theory based on Eqs. (\ref{b1})-(\ref{b3}) has been discussed by Reis (2003) who showed that, depending on the equation of state, the results may agree or disagree with the general relativistic approach. We shall leave this problem open since it is not our purpose here to investigate the validity of these equations in detail. Furthermore, our main results will concern the evolution of the cosmic fluid governed by the Friedmann equations. Since the Friedmann equations can be derived from the theory of general relativity, the validity of these equations is not questioned. Note that equations different from Eqs. (\ref{b1})-(\ref{b3}) have been considered by Pace {et al.} (2010) and Harko (2011). Their equations involve an additional term $\dot p {\bf u}/c^2$ in the Euler equation (\ref{b2}). However, this term apparently leads to equations for the cosmic evolution that are different from the Friedmann equations, so that this term will not be considered here.

\subsection{The cosmic evolution of a BEC universe}
\label{sec_cosmic}

We shall assume that the dark matter is a BEC with quartic self-interaction\footnote{In this part of the paper, we neglect the quantum pressure.} described by the barotropic equation of state (\ref{gpp8}). For convenience, we write this equation in the form
\begin{equation}
\label{b4}
p=kc^2\rho^2,
\end{equation}
where $k=2\pi a_s\hbar^2/m^3c^2$ is a constant that can be positive ($a_s>0$) or negative ($a_s<0$). In a first step, we concentrate on the dark matter component and neglect radiation, baryonic matter and dark energy (in particular we take $\Lambda=0$). These additional terms will be considered later. For the equation of state (\ref{b4}), the Friedmann equation (\ref{b5}) becomes
\begin{equation}
\label{cos1}
\frac{d\rho_b}{dt}+3\frac{\dot a}{a}\rho_b (1+k\rho_b)=0.
\end{equation}
This equation can be integrated into
\begin{equation}
\label{cos2}
\rho_b=\frac{A}{a^3-kA},
\end{equation}
where $A$ is a constant. If we make the physical requirement that equation (\ref{cos2}) has solutions for large $a$, then we must impose $A>0$. In that case,  $\rho_b\sim A/a^3$ for $a\rightarrow +\infty$ which returns Eq. (\ref{nc4}).  When $k>0$, which is the case previously considered by Harko (2011), the density exists only for $a>a_*=(kA)^{1/3}$. For $a\rightarrow a_*$,  $\rho_b\rightarrow +\infty$. When $k<0$, the density is defined for all $a$ and has a finite value $\rho_b=1/|k|$ when $a\rightarrow 0$.

Combining Eqs. (\ref{b6}) and (\ref{b4}), we obtain
\begin{equation}
\label{cos3}
\frac{\ddot a}{a}=-\frac{4\pi G\rho_b}{3}(1+3k\rho_b).
\end{equation}
When $k>0$, the universe is always decelerating ($\ddot a<0$). When $k<0$, the universe is accelerating ($\ddot a>0$) for $\rho_b>\rho_c\equiv 1/3|k|$ and decelerating ($\ddot a<0$) for $\rho_b<\rho_c$. Using Eq. (\ref{cos2}), this can be re-expressed in terms of the radius\footnote{We shall sometimes call the scale factor $a$ the ``radius'', having in mind the naive picture of a universe behaving like a homogeneous sphere of radius $a$. This is of course an abuse of language in the context of the theory of general relativity.}: the universe is accelerating for $a<a_c\equiv (2A|k|)^{1/3}$ and decelerating for $a>a_c$.

In order to determine the temporal  evolution of $a(t)$, we shall assume $\kappa=0$ (flat space) like in the Einstein-de Sitter universe.  Combining Eqs. (\ref{b7}) and (\ref{cos2}), we get
\begin{equation}
\label{cos4}
\dot a=\left (\frac{8\pi GA}{3}\right )^{1/2}\frac{a}{\sqrt{a^3-kA}}.
\end{equation}
We note that for $a\rightarrow +\infty$, this equation reduces to Eq. (\ref{nc8}) with Eq. (\ref{nc4}) and we recover the Einstein-de Sitter solution (\ref{nc9}). It is convenient to define $a_*=(|k|A)^{1/3}$ and introduce $R=a/a_*$. In that case, the density is given by
\begin{equation}
\label{cos7b}
\rho_b=\frac{1}{|k|}\frac{1}{R^3\mp 1},
\end{equation}
and Eq. (\ref{cos4}) can be rewritten
\begin{equation}
\label{cos5}
\dot R=\frac{KR}{\sqrt{R^3\mp 1}},
\end{equation}
where the upper sign $-$ corresponds to $k>0$ and the lower sign $+$ corresponds to $k<0$. We have defined the constant
\begin{equation}
\label{cos6}
K=\left (\frac{8\pi G}{3|k|}\right )^{1/2}=\left (\frac{4Gm^3c^2}{3|a_s|\hbar^2}\right )^{1/2}.
\end{equation}
Note that it depends on the mass $m$ of the bosons and on their scattering length $a_s$. Introducing the dimensionless parameter $\lambda=a_s/\lambda_c=a_smc/\hbar$ (Chavanis 2011) measuring the strength of the short-range interactions ($\lambda_c$ is the Compton wavelength of the bosons), we can rewrite Eq. (\ref{cos6}) in the form
\begin{equation}
\label{cos7}
K=\frac{2}{\sqrt{3}}\left (\frac{m}{M_P}\right )^2\frac{1}{\sqrt{|\lambda|}}t_P^{-1},
\end{equation}
where $M_P=(\hbar c/G)^{1/2}$ is the Planck mass, $l_P=(\hbar G/c^3)^{1/2}$ is the Planck length and $t_P=l_P/c$ is the Planck time.

\begin{figure}[!h]
\begin{center}
\includegraphics[clip,scale=0.3]{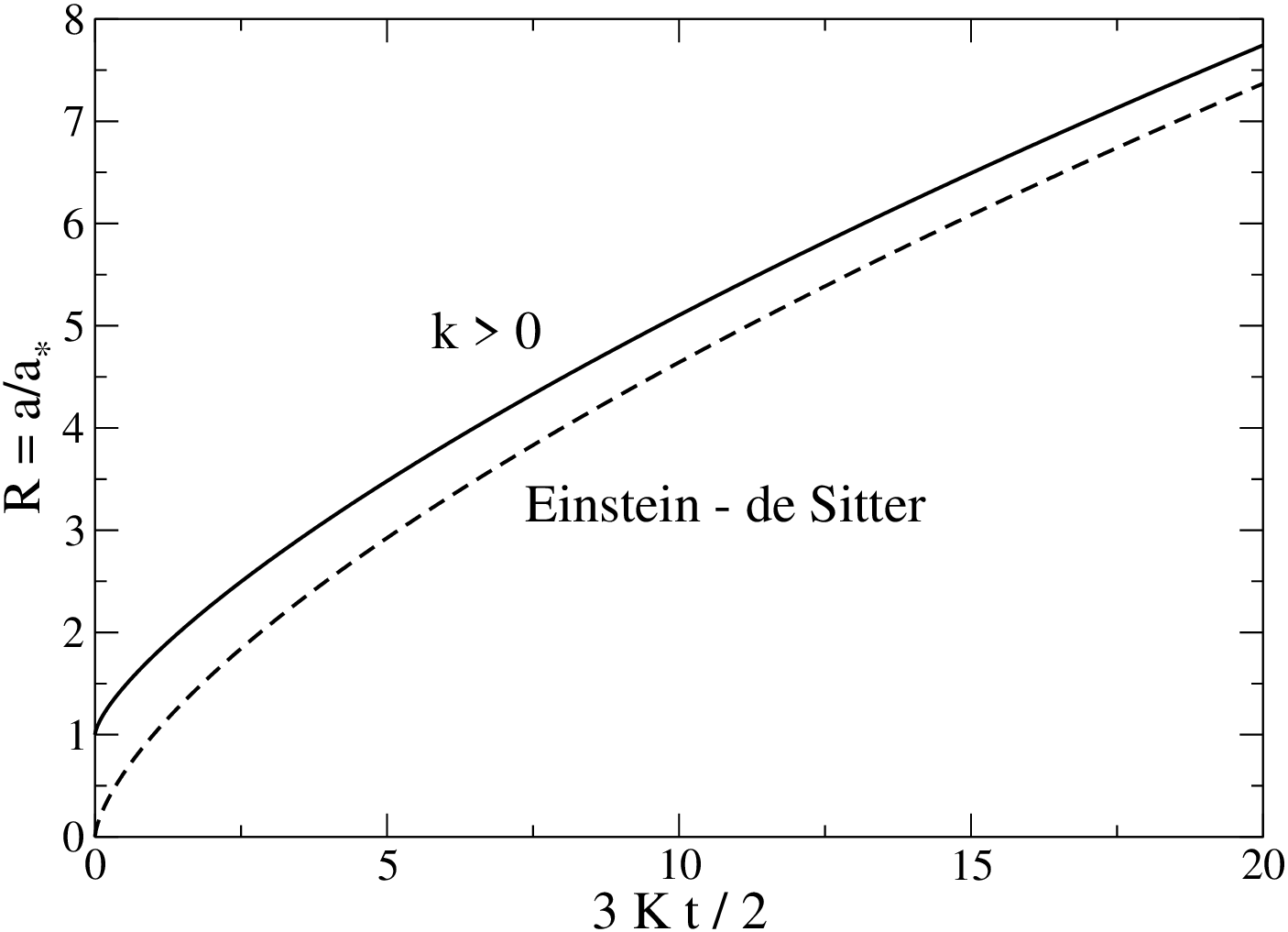}
\caption{Evolution of the scale factor in a BEC universe with $k>0$. The dashed line corresponds to the Einstein-de Sitter universe ($k=0$) that is reached asymptotically.}
\label{becscalingPLUS}
\end{center}
\end{figure}

For $k>0$, which is the case previously considered by Harko (2011), the solution of Eq. (\ref{cos5}) is
\begin{equation}
\label{cos8}
\sqrt{R^3-1}-\arctan \sqrt{R^3-1}=\frac{3}{2}Kt,
\end{equation}
where the constant of integration has been set equal to zero. In this model, the universe starts at a finite time $t=0$ (see Sec. \ref{sec_tl} for a revision of this statement) with a finite radius $R(0)=1$ and an infinite density $\rho_b(0)=\infty$. For $t\rightarrow 0$, $R\simeq 1+(3/4)^{1/3}(Kt)^{2/3}$. The universe is expanding, always decelerating, and asymptotically approaches the Einstein-de Sitter universe $R\sim (3Kt/2)^{2/3}$ (see Fig. \ref{becscalingPLUS}).

\begin{figure}[!h]
\begin{center}
\includegraphics[clip,scale=0.3]{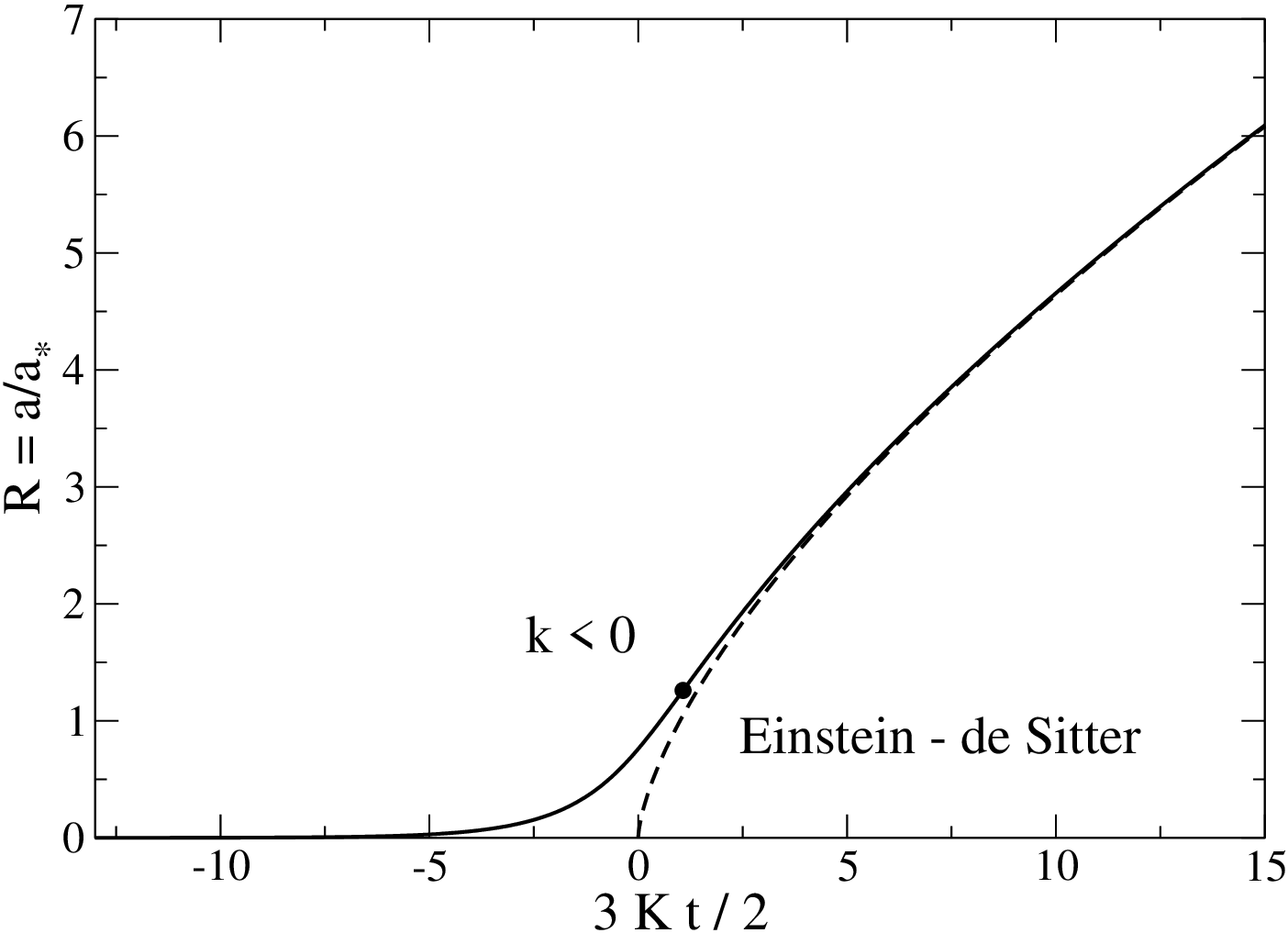}
\caption{Evolution of the scale factor in a BEC universe with $k<0$. The dashed line corresponds to the Einstein-de Sitter universe ($k=0$). We note that a BEC universe with $k<0$ converges more rapidly towards the EdS solution than a BEC universe with $k>0$.  The bullet (locating the inflexion point) corresponds to the time $t_c$ at which the universe starts to decelerate.}
\label{becscalingMOINS}
\end{center}
\end{figure}

For $k<0$, the solution of Eq. (\ref{cos5}) is
\begin{equation}
\label{cos9}
\sqrt{R^3+1}-\ln \left (\frac{1+\sqrt{R^3+1}}{R^{3/2}}\right )=\frac{3}{2}Kt,
\end{equation}
where the constant of integration has been set equal to zero. In this model, the universe starts from  $t\rightarrow -\infty$ with a vanishing radius $R(-\infty)=0$ and a finite density $\rho_b(-\infty)=1/|k|$. For $R<R_c\equiv 2^{1/3}\simeq 1.26$, the universe is accelerating and for $R>R_c\equiv 2^{1/3}$, it is decelerating. For $R\rightarrow +\infty$, we recover the classical Einstein-de Sitter universe $R\sim (3Kt/2)^{2/3}$ (see Fig. \ref{becscalingMOINS}). The change of regime between the phase of acceleration and the phase of deceleration manifests itself by an inflexion point ($\ddot R=0$). The time at which the universe starts decelerating is
\begin{equation}
\label{cos10}
\frac{3}{2}Kt_c= \sqrt{3}-\ln\left (\frac{1+\sqrt{3}}{\sqrt{2}}\right )\simeq 1.074.
\end{equation}
At $t=0$, $R(0)\simeq 0.7601$ and $\rho_b(0)=0.6948/|k|$. For $t\rightarrow -\infty$,
$R\rightarrow 0$ and the asymptotic expansion of
Eq. (\ref{cos9}) yields
\begin{equation}
\label{cos11}
R(t)\propto e^{Kt}.
\end{equation}
For $t\ll t_c$, the universe  expands exponentially rapidly, decelerates after $t_c$ and coincides with  the Einstein-de Sitter universe for $t\gg t_c$. In the accelerating phase, the rate of the expansion is given by Eq. (\ref{cos6}).

{\it Important remark:} in our mathematical description of the evolution of the cosmic fluid, we have extrapolated the solutions (\ref{cos8}) and (\ref{cos9}) far away in the past. In the model with $k>0$,
the universe emerges at a primordial time $t=0$ from a big-bang singularity where the density is infinite. In the model with $k<0$, the universe has always existed (up to $t\rightarrow -\infty$) and there is no big bang singularity at $t=0$. This is due to the negative pressure allowing for the presence of an inflexion point that changes the concavity of the curve $a(t)$.  Of course, this extrapolation to $t\rightarrow -\infty$ is not physically justified since we have ignored important effects like the radiation which dominates in the early universe. We shall see in Sec. \ref{sec_contribution} that the contribution of the radiation strongly alters the results at early times in the case $k<0$. Therefore, our description of the solution  (\ref{cos9}) must be considered only on a formal mathematical basis. Still it shows that one can construct cosmological models (solution to the Einstein equations) without primordial singularity.

\subsection{Other representations}
\label{sec_rep}

The previous representation shows that there exists a {\it single} universal curve $a(t)$ in each case  $a_s>0$, $a_s=0$ (EdS) and $a_s<0$, provided that the units of time and length are appropriately chosen. However, these units depend on $a_s$. Alternatively, it may be useful to choose units of time and length that are independent on $a_s$ and plot the curve $a(t)$ for different values of $a_s$. This is the representation that has been chosen by Harko (2011) for the case $k>0$ and that we shall discuss and generalize  in this section.

\begin{figure}[!h]
\begin{center}
\includegraphics[clip,scale=0.3]{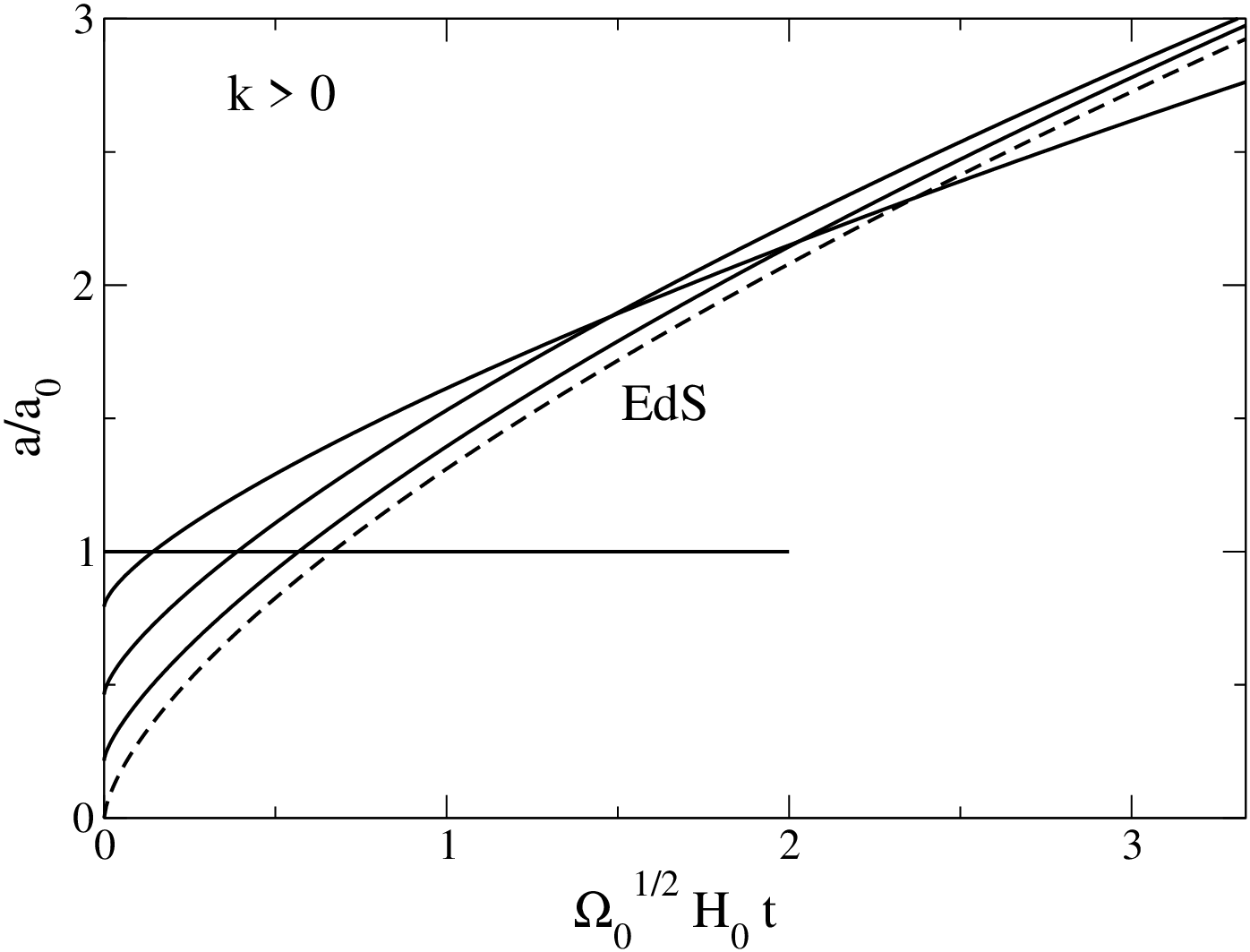}
\caption{Scale factor $a(t)$ as a function of time $t$ for a BEC universe with $k\ge 0$. The units of space and time are normalized by present-day quantities that do not depend on $k$. The different curves correspond to $q=0$ (EdS, dashed), $q=0.01$, $q=0.1$ and $q=0.5$. The straight line corresponds to $a/a_0=1$. Its intersection with the curve $a(t)$ defines the age of the universe.}
\label{becharkoPLUS}
\end{center}
\end{figure}

\begin{figure}[!h]
\begin{center}
\includegraphics[clip,scale=0.3]{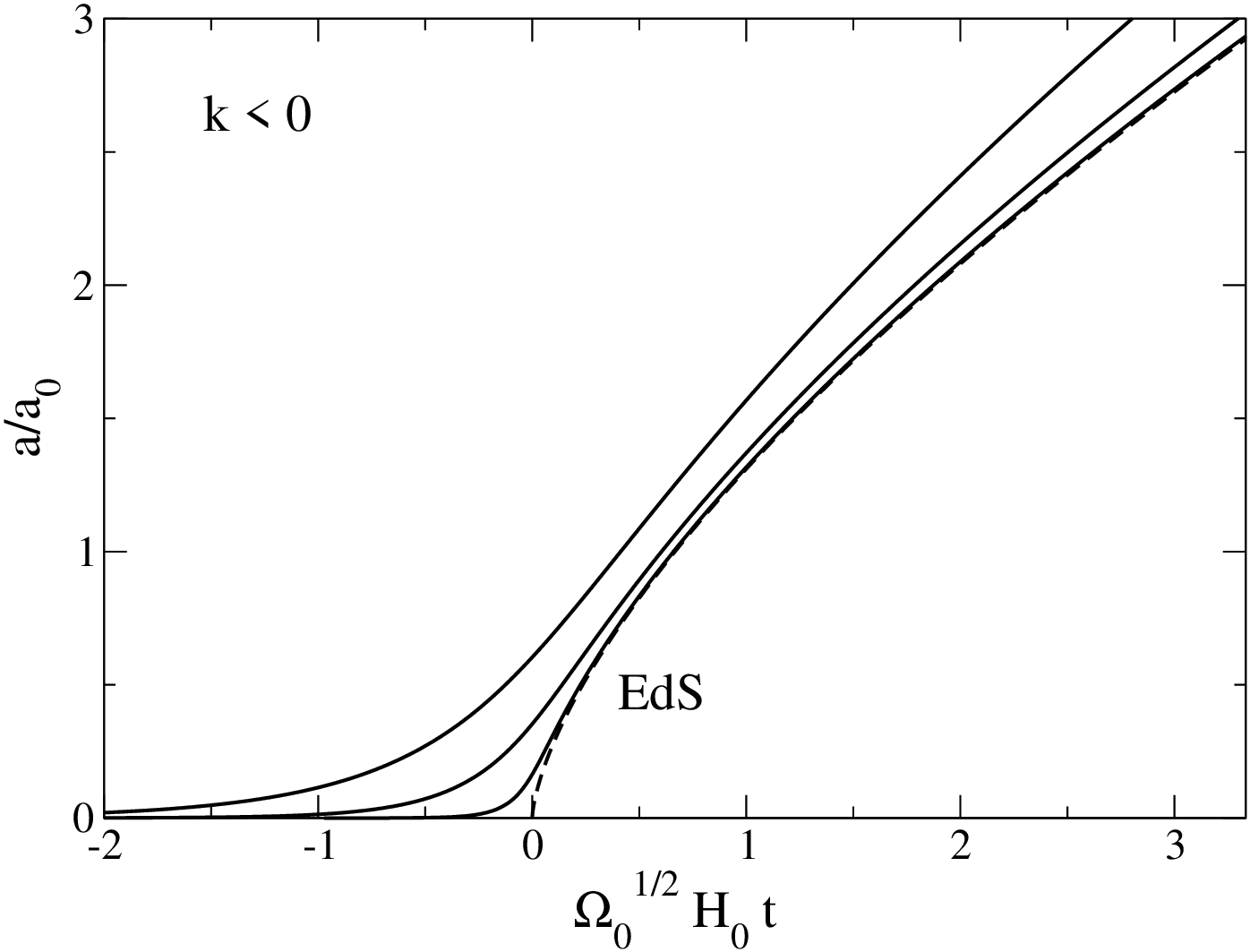}
\caption{Scale factor $a(t)$ as a function of time $t$ for a BEC universe with $k\le 0$. The units of space and time are normalized by present-day quantities that do not depend on $k$. The different curves correspond to $q=0$ (EdS, dashed), $q=-0.01$, $q=-0.1$ and $q=-0.5$.}
\label{becharkoMOINS}
\end{center}
\end{figure}

In Eq. (\ref{cos2}), the constant $A$ can be determined by the present-day density $\rho_b=\rho_0$ and the corresponding scale factor $a=a_0$. Writing $\rho_0=A/(a_0^3-kA)$, we obtain
\begin{eqnarray}
\label{rep1}
A=\frac{\rho_0 a_0^3}{1+k\rho_0}.
\end{eqnarray}
We note that $\rho_0<1/|k|$ when $k<0$ as we have already indicated.  The relation (\ref{cos2}) between the density and the scale factor can be rewritten
\begin{eqnarray}
\label{rep2}
\rho_b=\frac{q}{k}\frac{1}{(\frac{a}{a_0})^3-q},
\end{eqnarray}
where we have introduced the dimensionless parameter
\begin{eqnarray}
\label{rep3}
q=\frac{k\rho_0}{1+k\rho_0}.
\end{eqnarray}
We note that $q$ is positive for $k>0$ and negative for $k<0$. Furthermore, $q=0$ for $k=0$,  $q\rightarrow 1$ for $k\rightarrow +\infty$ and $q\rightarrow -\infty$ for $k\rightarrow -1/\rho_0$.
Using these results, the differential equation (\ref{cos4}) giving the evolution of the scale factor with the time can be rewritten
\begin{eqnarray}
\label{rep4}
\dot a=\left (\frac{8\pi G q}{3k}\right )^{1/2}\frac{a}{\sqrt{(\frac{a}{a_0})^3-q}}.
\end{eqnarray}
From Eq. (\ref{b7}), we can define the present-day critical density by $(\rho_c)_0=3H_0^2/8\pi G$ where $H_0$ is the present-day Hubble constant. If $\kappa=0$, which seems to be the case for our universe,  $(\rho_c)_0$ represents  the total density including baryonic matter, radiation, dark matter and dark energy. It is customary to introduce the present-day density parameter $\Omega_0=\rho_0/(\rho_c)_0$. Then, we can write
\begin{eqnarray}
\label{rep5}
\left (\frac{8\pi G q}{3k}\right )^{1/2}=\frac{H_0\sqrt{\Omega_0}}{\sqrt{1+k\rho_0}}.
\end{eqnarray}
Since the denominator depends on $k$, it must be expressed in terms of $q$ (which is our control parameter). Using Eq. (\ref{rep3}), we obtain
\begin{eqnarray}
\label{rep6}
k\rho_0=\frac{q}{1-q}.
\end{eqnarray}
so that  Eq. (\ref{rep4}) can be rewritten
\begin{eqnarray}
\label{rep7}
\dot a=H_0\sqrt{\Omega_0}(1-q)^{1/2}\frac{a}{\sqrt{(\frac{a}{a_0})^3-q}}.
\end{eqnarray}
The density is
\begin{eqnarray}
\label{rep2b}
\rho_b=\rho_0\frac{1-q}{(\frac{a}{a_0})^3-q},
\end{eqnarray}
We can make the connection with the notations of Sec. \ref{sec_cosmic} by setting
\begin{eqnarray}
\label{rep8}
a_*=a_0 |q|^{1/3},\qquad K=H_0\sqrt{\Omega_0}\left (\frac{1-q}{|q|}\right )^{1/2}.
\end{eqnarray}
In that case, Eq. (\ref{rep7}) takes the form of Eq. (\ref{cos5}) and its solutions are given by Eqs. (\ref{cos8}) and (\ref{cos9}). Returning to the notations of this section, we find for $0\le q<1$ that
\begin{eqnarray}
\label{rep9}
\frac{1}{\sqrt{1-q}}\biggl\lbrace \sqrt{(\frac{a}{a_0})^3-q}-\sqrt{q}\arctan \sqrt{\frac{(\frac{a}{a_0})^3-q}{q}}\biggr\rbrace\nonumber\\
=\frac{3}{2}\sqrt{\Omega_0}H_0t,
\end{eqnarray}
which was previously obtained by Harko (2011). On the other hand, for $q\le 0$, we obtain
\begin{eqnarray}
\label{rep10}
\frac{1}{\sqrt{1-q}}\biggl\lbrace \sqrt{(\frac{a}{a_0})^3-q}-\sqrt{|q|}\ln \left (\frac{\sqrt{|q|}+ \sqrt{(\frac{a}{a_0})^3-q}}{(\frac{a}{a_0})^{3/2}}\right )\biggr\rbrace\nonumber\\
=\frac{3}{2}\sqrt{\Omega_0}H_0 t.\qquad
\end{eqnarray}
Finally, for $q=0$, we recover the Einstein-de Sitter model
\begin{eqnarray}
\label{rep11}
\frac{a}{a_0}=\left (\frac{3}{2}\sqrt{\Omega_0}\right )^{2/3}(H_0t)^{2/3}.
\end{eqnarray}
The curves giving the scale factor $a(t)$ as a function of time for different values of $q$ are plotted in Figs. \ref{becharkoPLUS} and \ref{becharkoMOINS}. The interest of this representation is that it shows how the BEC models (\ref{rep9}) and (\ref{rep10}) approach the Einstein-de Sitter model (\ref{rep11}) as $q\rightarrow 0$.  On the other hand, for larger values of $q$ (positive or negative), there can be substantial differences between a BEC universe and the classical pressureless Einstein-de Sitter universe.

To see that, let us consider the asymptotic behaviors of Eqs. (\ref{rep9}) and (\ref{rep10}). For $t=0$, we find that
\begin{eqnarray}
\label{rep16}
\frac{a}{a_0}=R_{\pm}|q|^{1/3},
\end{eqnarray}
with $R_+=1$ for $k>0$ and $R_{-}\simeq 0.7601$ for $k<0$. For $t\rightarrow +\infty$, we get
\begin{eqnarray}
\label{rep12}
\frac{a}{a_0}\sim \left (\frac{3}{2}\sqrt{\Omega_0}\right )^{2/3}(1-q)^{1/3}(H_0t)^{2/3}.
\end{eqnarray}
For $0<q<1$, the scale factor is asymptotically smaller than in an EdS universe (corresponding to $q=0$). Since a BEC universe with $k>0$ initially starts with a radius $a(0)>0$, this implies that the curves must cross each other at some point (such a crossing is shown in Fig. \ref{becharkoPLUS} for $q=0.5$). Of course, the crossing point occurs at larger and larger times as $q\rightarrow 0$. For $q<0$, the scale factor is always larger than in an EdS universe.

For $q<0$, the inflexion point indicating when the universe starts decelerating is located at
\begin{eqnarray}
\label{rep17}
\frac{a_c}{a_0}=2^{1/3}|q|^{1/3},
\end{eqnarray}
\begin{eqnarray}
\label{rep17b}
\frac{3}{2}\sqrt{\Omega_0}H_0t_c=\left\lbrack \sqrt{3}-\ln\left (\frac{1+\sqrt{3}}{\sqrt{2}}\right )\right\rbrack \left (\frac{|q|}{1-q}\right )^{1/2}.\nonumber\\
\end{eqnarray}
For $q\rightarrow -\infty$, $a_c\rightarrow +\infty$ and $\frac{3}{2}\sqrt{\Omega_0}H_0t_c\rightarrow 1.074$. On the other hand for $t\rightarrow -\infty$, the scale factor behaves like
\begin{eqnarray}
\label{rep18}
\frac{a}{a_0}\propto e^{\sqrt{\frac{1+|q|}{|q|}\Omega_0}H_0t}.
\end{eqnarray}

\begin{figure}[!h]
\begin{center}
\includegraphics[clip,scale=0.3]{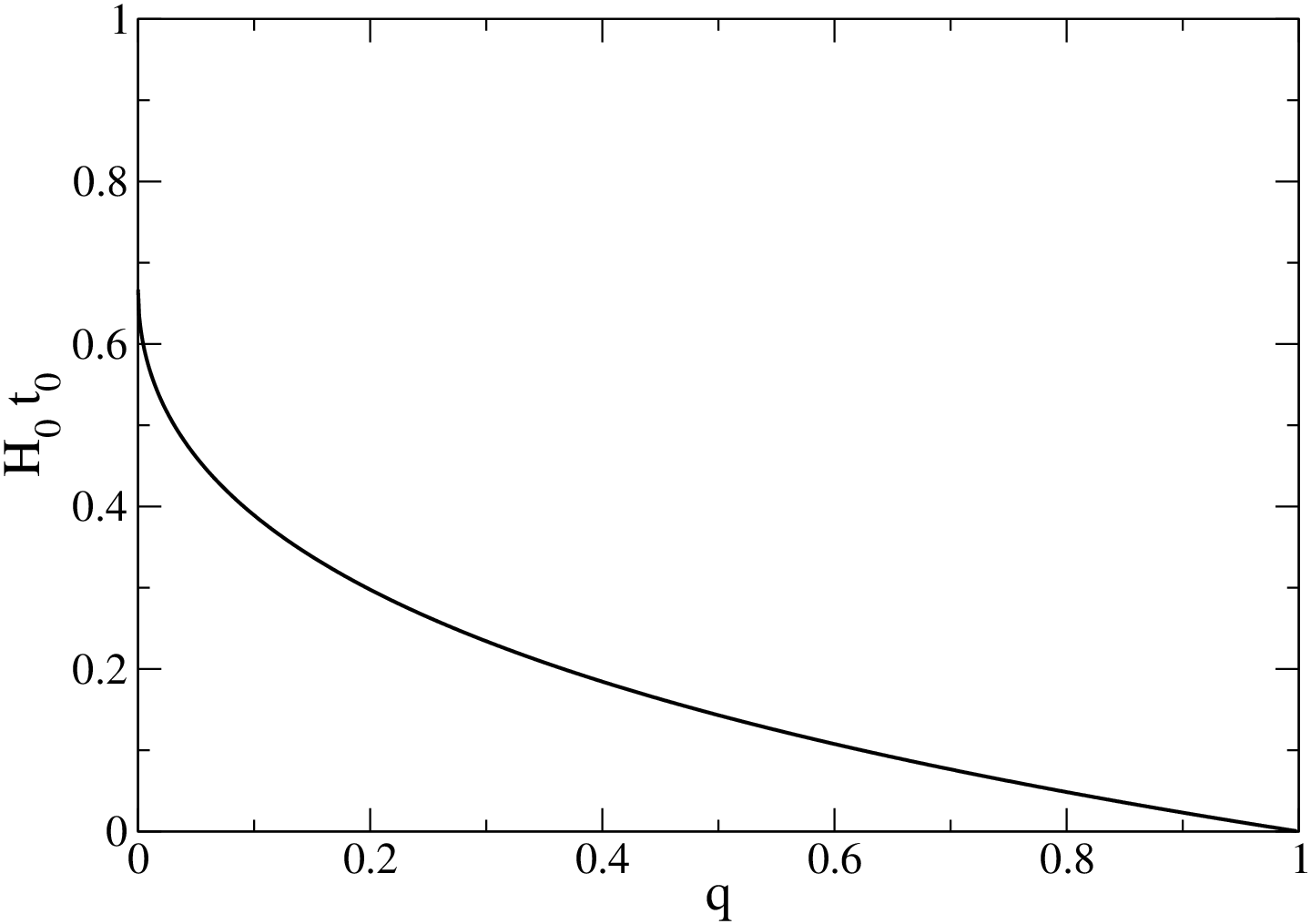}
\caption{Age of a BEC universe as a function of $q\ge 0$. The Einstein-de Sitter model corresponds to $q=0$. }
\label{age}
\end{center}
\end{figure}

Let us naively assume that all the content of the universe is in the
form of BEC dark matter and take $\Omega_0=1$. The age of a BEC
universe corresponds to the time $t_0$ at which $a/a_0=1$.  For $0\le q<1$,
\begin{eqnarray}
\label{rep13}
H_0 t_0=\frac{2}{3}\left\lbrace 1-\sqrt{\frac{q}{1-q}}\arctan \sqrt{\frac{1-q}{q}}\right\rbrace,
\end{eqnarray}
and for $q\le 0$,
\begin{eqnarray}
\label{rep13b}
H_0 t_0=\frac{2}{3}\biggl\lbrace 1-\sqrt{\frac{-q}{1-q}}\ln \left ({\sqrt{-q}+ \sqrt{1-q}}\right )\biggr\rbrace.
\end{eqnarray}
The age of the EdS universe ($q=0$) is
\begin{eqnarray}
\label{rep15}
H_0 t_0=\frac{2}{3}.
\end{eqnarray}
As can be seen in Fig. \ref{age}, the age of a BEC universe with $q\neq 0$ is smaller than the age of the EdS universe ($q=0$).

\subsection{Contribution of radiation, baryons and dark energy}
\label{sec_contribution}

Several observational results indicate that the universe is flat ($\kappa=0$) and that its present-day expansion is accelerating ($\ddot a >0$). Furthermore, the estimated age of the universe is
$t_0\sim 13.75\, {\rm Gyrs}$ which is about  $H_0^{-1}$. If we assume that the universe is made only of a pressureless
fluid and if we take $\Lambda=0$, we are led to the EdS
universe. However, this model leads to a decelerating expansion and
predicts a too small age of the universe $t_0=2/3H_0\sim 9.3\, {\rm Gyrs}$. If we assume that dark
matter is a BEC, we again find that the present-day universe is
decelerating. Furthermore, its predicted age is even smaller than in
the EdS universe (see Sec. \ref{sec_rep}). Therefore, we must invoke some form of dark energy to understand
the present-day acceleration of the universe and its age. In the
standard $\Lambda$CDM model, dark energy is due to the cosmological
constant. On the other hand, to be complete, we must include the contribution of dark matter, baryonic matter and radiation. In that case, Eq. (\ref{rep7}) is replaced by
\begin{eqnarray}
\label{rep19}
\frac{\dot a}{a}=H_0\sqrt{\frac{\Omega_{B,0}}{(a/a_0)^3}+\frac{\Omega_{rad,0}}{(a/a_0)^4}
+\frac{\Omega_{DM,0}(1-q)}{(a/a_0)^3-q}+\Omega_{\Lambda}},\nonumber\\
\end{eqnarray}
where $\Omega_{B,0}$,  $\Omega_{rad,0}$, $\Omega_{DM,0}$ and $\Omega_{\Lambda}$ are the present-day values of the density parameters of the baryonic matter, radiation, dark matter and dark energy. We have assumed that dark matter is in the form of BEC. The case of a pressureless dark matter is recovered for $q=0$. Following Harko (2011), we have adopted the numerical values $\Omega_{B,0}=0.0456$,  $\Omega_{rad,0}=8.24\, 10^{-5}$, $\Omega_{DM,0}=0.228$ and $\Omega_{\Lambda}=0.726$ (Hinshaw et al. 2009).

\begin{figure}[!h]
\begin{center}
\includegraphics[clip,scale=0.3]{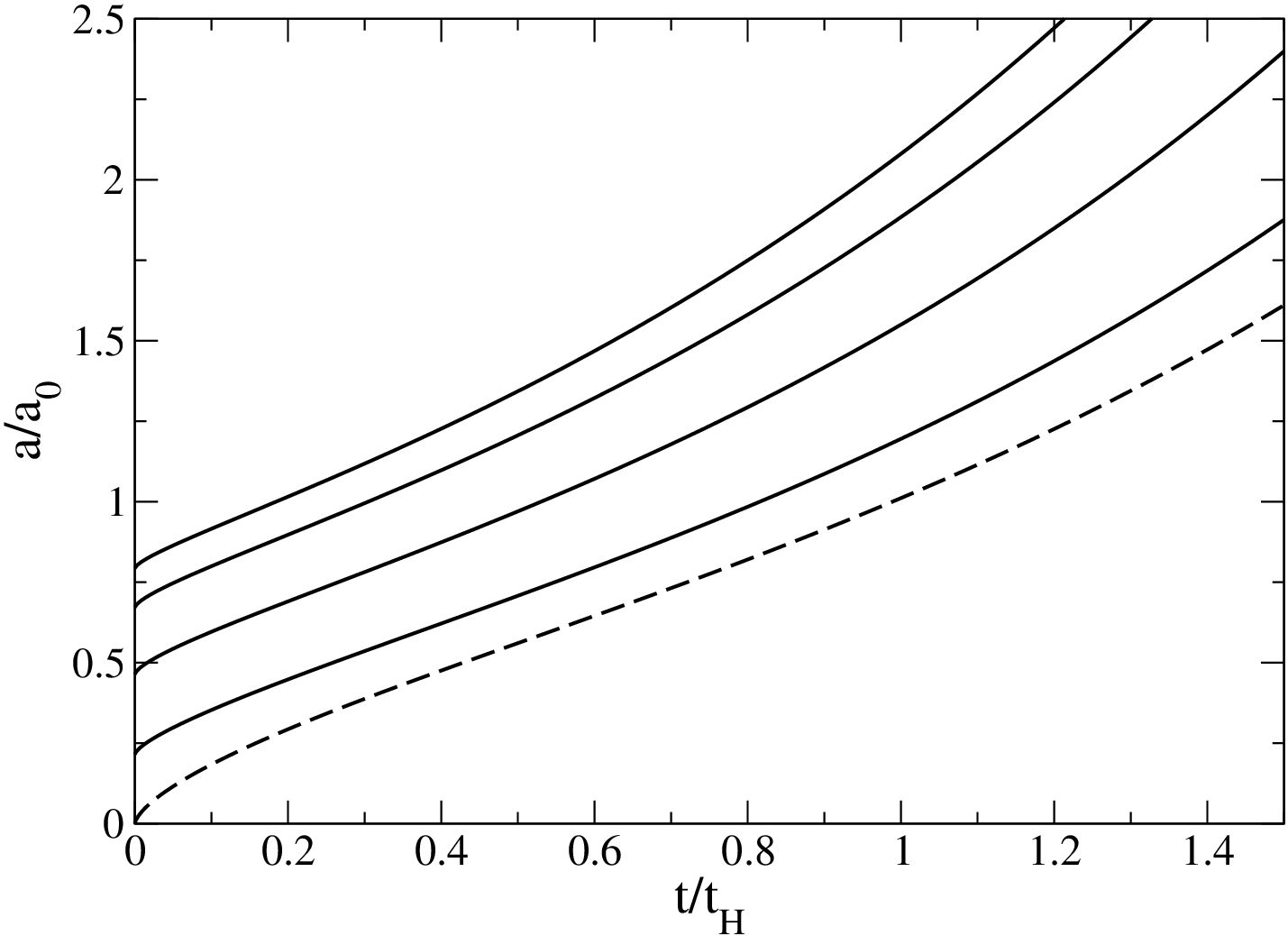}
\caption{Scale factor $a(t)$ as a function of time $t$ for a universe filled with baryonic matter, radiation, BEC dark matter with $k\ge 0$, and dark energy. The different curves correspond to $q=0$ (pressureless dark matter, dashed), $q=0.01$, $q=0.1$, $q=0.303471$ and $q=0.5$. }
\label{darkPOS}
\end{center}
\end{figure}

\begin{figure}[!h]
\begin{center}
\includegraphics[clip,scale=0.3]{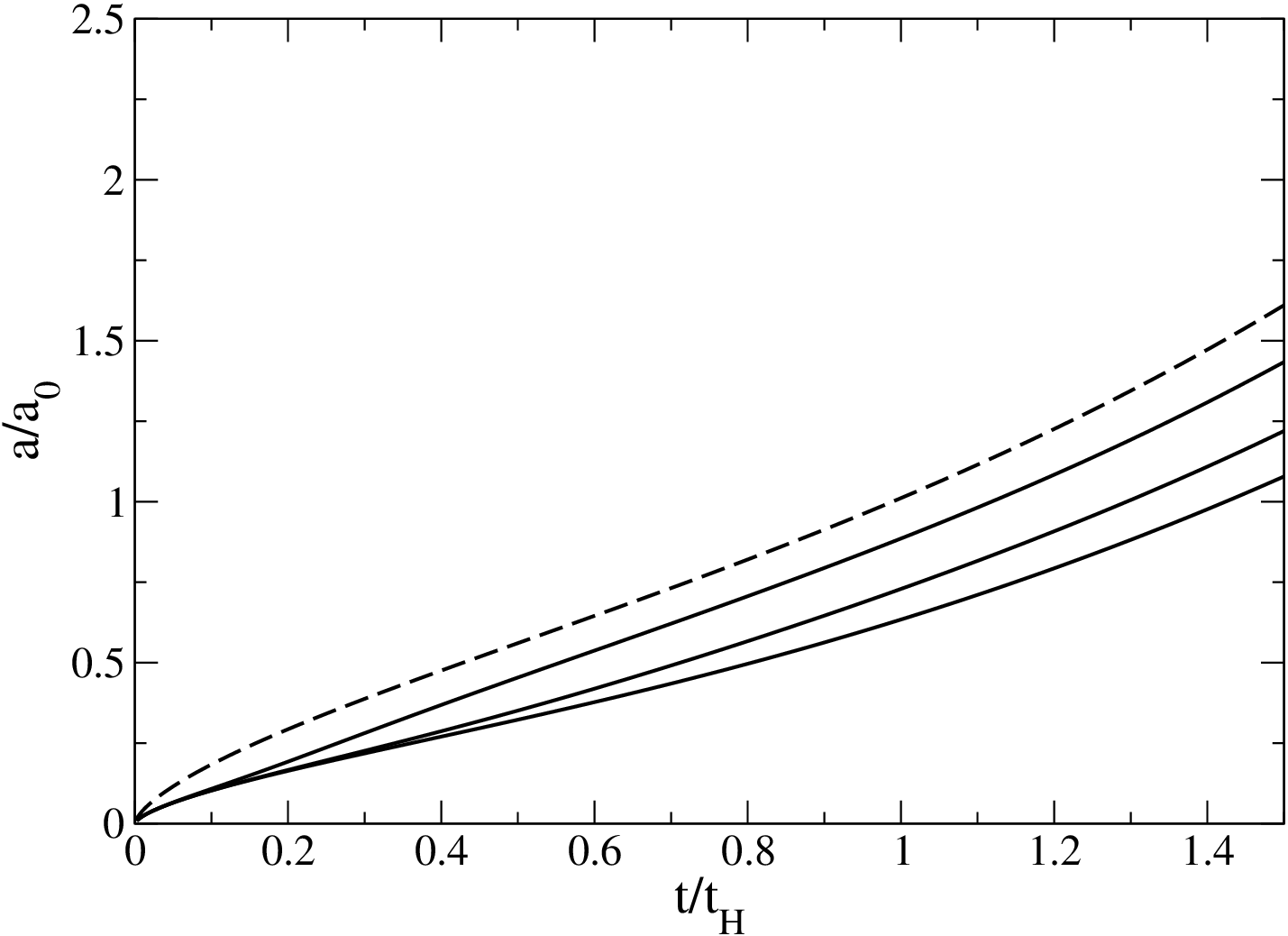}
\caption{Scale factor $a(t)$ as a function of time $t$ for a universe filled with baryonic matter, radiation, BEC dark matter with $k\le 0$, and dark energy. The different curves correspond to $q=0$ (pressureless dark matter, dashed), $q=-0.01$, $q=-0.1$ and $q=-0.5$. }
\label{darkNEG}
\end{center}
\end{figure}

The evolution of the scale factor $a(t)$ is represented in Figs. \ref{darkPOS} and \ref{darkNEG} for $k>0$ and $k<0$ respectively. In both cases, the universe is initially decelerating but finally enters in a phase of acceleration due to the effect of dark energy (cosmological constant). In the case $k>0$, previously treated by Harko (2011), the scale factor increases more rapidly than in the standard model ($k=0$). Furthermore, the density of dark matter becomes infinite at a finite radius so that radiation does not dominate in the early universe. Accordingly, the universe starts from $a(0)>0$ at $t=0$. In the case $k<0$, the scale factor increases less rapidly than in the standard model ($k=0$). Furthermore, the density of dark matter does not diverges when $a\rightarrow 0$ so that radiation dominates in the early universe. Accordingly, the universe starts from $a(0)=0$ with an infinite (radiation) density.

\subsection{Theory of linear perturbations}
\label{sec_tl}

The theory of linear perturbations based on the Newtonian equations with pressure (\ref{b1})-(\ref{b3}) has been performed by Reis (2003) in the general case, so that we shall directly apply  his results to the present situation. Note that Harko (2011) considers other equations and obtains different results. Nevertheless, our main conclusions will be the same. For an arbitrary equation of state, the evolution of the density contrast is given by (Reis, 2003):
\begin{eqnarray}
\label{tl1}
\ddot\delta-H\left\lbrack 3 (2w-c_{eff}^2-c_s^2)-2\right\rbrack\dot\delta\nonumber\\
+3H^2\biggl\lbrack \frac{1}{H}\dot{c_{eff}^2}
+(3c_s^2-6w-1)c_{eff}^2\nonumber\\
+\frac{3w^2}{2}-4w-\frac{1}{2}
+3c_s^2\biggr\rbrack \delta-\frac{c_{eff}^2c^2}{a^2}\Delta\delta=0,
\end{eqnarray}
where $H=\dot a/a$ is the Hubble constant, $w=p_b/\rho_b c^2$, $c_{eff}^2=(\delta p/\delta \rho)/c^2$ and $c_s^2=p_b'(\rho_b)/c^2$. These last quantities represent the velocity of sound normalized by the velocity of light. For a barotropic fluid, $c_{eff}^2=c_s^2=p'(\rho_b)/c^2$. On the other hand, from the Friedmann equations (\ref{b5})-(\ref{b7}), it is easy to establish the relation
\begin{eqnarray}
\label{tl2}
\frac{\dot w}{1+w}=-3H(c_s^2-w).
\end{eqnarray}

We shall now focus on the equation of state (\ref{b4}). For this equation of state, we obtain $w=k\rho_b$ and $c_{eff}^2=c_s^2=2w$.  In that case, Eq. (\ref{tl1}) reduces to
\begin{equation}
\label{tl3}
\ddot\delta+2H (3w+1)\dot\delta-3H^2 \left (\frac{9w^2}{2}+6w+\frac{1}{2}\right ) \delta
-\frac{2wc^2}{a^2}\Delta\delta=0,
\end{equation}
where we have used $\dot w/H=-3w(1+w)$ according to Eq. (\ref{tl2}).  Measuring the evolution in terms of $a$ rather than in terms of $t$, and using the Friedmann equations (\ref{b6}) and (\ref{b7}), we can rewrite Eq. (\ref{tl3}) in the form
\begin{equation}
\label{tl4}
a^2\frac{d^2 \delta}{da^2}+\frac{3}{2}(1+3w) a \frac{d\delta}{da}-\frac{3}{2} (9w^2+12w+1)\delta -\frac{2wc^2}{H^2a^2}\Delta\delta=0.
\end{equation}
Finally, according to Eq. (\ref{cos7b}), we have
\begin{eqnarray}
\label{tl5}
w=\frac{\pm 1}{R^3\mp 1},
\end{eqnarray}
where the upper sign corresponds to $k>0$ and the lower sign corresponds to $k<0$. If, for simplicity, we ignore the Laplacian term in Eq. (\ref{tl4}), we obtain
\begin{eqnarray}
\label{tl6}
R^2\frac{d^2 \delta}{dR^2}+\frac{3}{2}(1+3w) R \frac{d\delta}{dR}-\frac{3}{2} (9w^2+12w+1)\delta=0.\nonumber\\
\end{eqnarray}
For $R\rightarrow +\infty$, $w\rightarrow 0$ and Eq. (\ref{tl6}) reduces to
\begin{eqnarray}
\label{tl7}
R^2\frac{d^2 \delta}{dR^2}+\frac{3}{2} R \frac{d\delta}{dR}-\frac{3}{2}\delta=0.
\end{eqnarray}
This returns the usual equation (\ref{eds2}) of a pressureless fluid in the Einstein-de Sitter universe. Its solutions are given by Eq. (\ref{eds3}).

Let us now consider the case of small $R$. For $k>0$, we have seen that the universe exists only for $R\ge 1$. Now, for $R\rightarrow 1$,  $w\rightarrow +\infty$ so that the differential equation (\ref{tl6}) is ill-defined. This means that we must start the perturbation analysis at $t=t_i>0$. Harko (2011) remarks that the requirement $c_{eff}=c_s<1$ (meaning that the velocity of sound must be smaller than the velocity of light) implies $R>3^{1/3}$ and $w<1/2$. Therefore, we shall assume that the universe starts at $t_*$ such that $R(t_*)=3^{1/3}$. According to Eq. (\ref{cos8}), we have
\begin{equation}
\label{cos10edf}
\frac{3}{2}Kt_*= \sqrt{2}-\arctan\sqrt{2}\simeq 0.4589.
\end{equation}
For $t\rightarrow t_*$, Eq. (\ref{tl6}) can be approximated by
\begin{eqnarray}
\label{tl6bis}
R^2\frac{d^2 \delta}{dR^2}+\frac{15}{4} R \frac{d\delta}{dR}-\frac{111}{8}\delta=0,
\end{eqnarray}
and its solutions are
\begin{eqnarray}
\label{tl11bis}
\delta \propto R^{\frac{-11+\sqrt{1009}}{8}},\qquad \delta \propto R^{\frac{-11-\sqrt{1009}}{8}}.
\end{eqnarray}
The growth of the perturbations is faster than in a cold EdS universe ($a^{2.6}$ instead of $a$). On the other hand, for $k<0$, the universe starts at $R=0$ for $t\rightarrow -\infty$ implying $w=-1$.
In that case,  Eq. (\ref{tl6}) can be approximated by
\begin{eqnarray}
\label{tl10}
R^2\frac{d^2 \delta}{dR^2}-3 R \frac{d\delta}{dR}+3\delta=0,
\end{eqnarray}
and its solutions are
\begin{eqnarray}
\label{tl11}
\delta \propto R^{3},\qquad \delta \propto R.
\end{eqnarray}
Contrary to a classical cold universe, the two solutions of the differential equation (\ref{tl10}) are growing. Furthermore, the growth of the perturbations is faster than in an EdS universe ($a^3$ instead of $a$). Combining Eq. (\ref{tl11}) with Eq. (\ref{cos11}), we get
\begin{eqnarray}
\label{tl12}
\delta(t) \propto e^{3Kt},\qquad \delta(t) \propto e^{Kt},
\end{eqnarray}
for $t\rightarrow -\infty$.

In conclusion, these simple estimates indicate that the growth of the
perturbations is faster in a BEC universe than in a pressureless
universe. This conclusion was previously reached by Harko (2011) for
$k>0$ based on different equations. Our approach confirms this
conclusion and extends it to $k<0$ (It may be recalled that the
Laplacian term in Eq. (\ref{tl3}) has been neglected in our simple
analysis. Now, we have seen in Sec. \ref{sec_s} that it is precisely
this term that leads to an increase of the maximum growth rate in the
(static) Jeans problem when $k<0$. Therefore, the inclusion of this
term in our analysis should enhance the growth of perturbations in the
case of attractive self-interaction).

\section{Dark fluid with a generalized equation of state}
\label{sec_dark}

\subsection{Linear equation of state}
\label{sec_leos}

One possibility to understand the present-day acceleration of the
expansion of the universe is to invoke a form of dark energy arising
from a non-zero value of the cosmological constant $\Lambda$ (see
Sec. \ref{sec_contribution}). However, the physical meaning of
$\Lambda$ is not clearly understood. A major problem is that most
quantum field theories predict a huge cosmological constant from the
energy of the quantum vacuum, more than $100$ order of magnitude too
large. Therefore, other approaches have been developed to understand
the phase of acceleration without invoking the cosmological
constant. One possibility is to consider a ``dark fluid'' with a
negative pressure. In this context, many workers (see the review of
Peebles \& Ratra 2003) have considered a linear equation of state
\begin{equation}
\label{dark0}
p=\alpha \rho c^2,
\end{equation}
with $-1\le \alpha\le 1$. For this equation of state, the Friedmann equations  (\ref{b5}), (\ref{b6}) and (\ref{b7}) with $\kappa=\Lambda=0$  reduce to
\begin{equation}
\label{dark1}
\frac{d\rho_b}{dt}+3\frac{\dot a}{a}(1+\alpha)\rho_b=0,
\end{equation}
\begin{equation}
\label{dark2}
\frac{\ddot a}{a}=-\frac{4\pi G}{3} (1+3\alpha)\rho_{b},\qquad \left (\frac{\dot a}{a}\right )^2=\frac{8\pi G}{3}\rho_b.
\end{equation}
Equation (\ref{dark1}) leads to the relation $\rho_{b}a^{3(1+\alpha)}\sim 1$. We note that the effect of a cosmological constant $\Lambda$ is equivalent to a fluid with an equation of state $p=-\rho c^2$ ($\alpha=-1$)  since, in that case, $\rho_b$ is constant and can be written $\rho_b=\rho_\Lambda\equiv \Lambda/8\pi G$. This equation of state   leads to an exponential growth of the scale factor  
\begin{equation}
\label{dark3bb}
a=a_0 e^{\sqrt{\Lambda/3}t}.
\end{equation}
On the other hand, for $-1<\alpha\le 1$, Eqs. (\ref{dark1}) and (\ref{dark2}) generate a model of the form
\begin{equation}
\label{dark3}
a\propto t^{2/\lbrack 3(1+\alpha)\rbrack},\qquad H=\frac{\dot a}{a}=\frac{2}{3(1+\alpha)t},
\end{equation}
\begin{equation}
\label{dark4}
\rho_b=\frac{1}{6\pi G(1+\alpha)^2 t^2}.
\end{equation}
According to Eq. (\ref{dark2}-a), this universe is accelerating for $-1<
\alpha<\alpha_{c}\equiv -1/3$ and decelerating for
$\alpha>\alpha_{c}$. For $\alpha=\alpha_c$, the scale factor increases
linearly with time ($a\propto t$). The EdS universe corresponds to
$\alpha=0$. For the general model (\ref{dark3})-(\ref{dark4}), the age
of the universe is $t_0=2/\lbrack 3(1+\alpha)H_{0}\rbrack$ where
$H_0=2.273\, 10^{-18}\, {\rm s}^{-1}$ is the present-day value of the
Hubble constant. We note that for the critical value $\alpha_c=-1/3$,
the age of the universe is $t_0=H_0^{-1}=13.95\, {\rm Gyrs}$ which is
close to the value $13.75\, {\rm Gyrs}$ predicted by the
$\Lambda$CDM model.  However, the deceleration parameter $Q=-{\ddot
a}a/{\dot a}^2=(1+3\alpha)/2$ vanishes for $\alpha=-1/3$ while its
present-day value is $\sim -0.5$. We also note that the linear equation
of state (\ref{dark0}) does not allow for a transition between a phase
of deceleration and a phase of acceleration, while such a transition
is likely in our universe.  It may be therefore interesting to
generalize this model.

\subsection{Generalized equation of state}
\label{sec_geos}

We consider a generalized equation of state of the form
\begin{equation}
\label{dark5}
p=(\alpha \rho+k\rho^2) c^2,
\end{equation}
with $-1\le \alpha\le 1$ and $k$ positive or negative (the case $\alpha=-1$ is specifically treated in Appendix \ref{sec_eosg}). In this approach, a single ``dark fluid'' combines the properties of a  BEC dark matter described by the equation of state (\ref{b4}) and of a dark energy described by the equation of state (\ref{dark0}). We note that the equation of state of the BEC dark matter ($\propto \rho^2$) dominates in the early universe where the density is high while the dark energy ($\propto \rho$) dominates in the present-day universe where the density is low. The equation of state of the BEC also dominates at the scale of dark matter halos. This makes the study of this equation of state interesting. Another nice feature of this equation of state is that it admits fully analytical solutions\footnote{Eq. (\ref{dark5}) can be written as $p_b=w(t)\rho_b c^2$ where $w(t)$ is a function of time. Many authors have considered an equation of state of that form with some prescribed function $w(t)$. In our model, $w(t)$ is not an {\it ad hoc} function but it is explicitly given by $w(t)=\alpha+k\rho_b(t)$.}.

For the equation of state (\ref{dark5}), the Friedmann equation (\ref{b5}) becomes
\begin{equation}
\label{dark6}
\frac{d\rho_b}{dt}+3\frac{\dot a}{a}\rho_b (1+\alpha+k\rho_b)=0.
\end{equation}
This equation can be integrated into
\begin{equation}
\label{dark7}
\rho_b=\frac{A(1+\alpha)}{a^{3(1+\alpha)}-kA},
\end{equation}
where $A>0$ is a constant. For $a\rightarrow +\infty$, $\rho_b\sim
A(1+\alpha)/a^{3(1+\alpha)}$.  When $k>0$, the density
exists only for $a>a_*=(kA)^{1/\lbrack 3(1+\alpha)\rbrack}$. For $a\rightarrow a_*$,
$\rho_b\rightarrow +\infty$. When $k<0$, the density is defined for
all $a$ and has a finite value $\rho_b=(1+\alpha)/|k|$ when $a\rightarrow 0$.

Combining Eqs. (\ref{b6}) and (\ref{dark5}), and taking $\Lambda=0$, we obtain
\begin{equation}
\label{dark8}
\frac{\ddot a}{a}=-\frac{4\pi G\rho_b}{3}(1+3\alpha+3k\rho_b).
\end{equation}
Let us define a critical density and a critical scale factor
\begin{equation}
\label{dark9}
\rho_c= -\frac{1+3\alpha}{3k},\qquad a_c=\left (\frac{-2kA}{1+3\alpha}\right )^{\frac{1}{3(1+\alpha)}},
\end{equation}
corresponding to a possible inflexion point ($\ddot a=0$) in the curve
$a(t)$. When $k>0$ and $\alpha\ge -1/3$, the universe is always
decelerating ($\ddot a<0$). When $k>0$ and $\alpha<-1/3$, the universe
is decelerating for $\rho_b>\rho_c$ (i.e. $a<a_c$) and accelerating
for $\rho_b<\rho_c$ (i.e. $a>a_c$). When $k<0$ and $\alpha\le -1/3$,
the universe is always accelerating ($\ddot a>0$). When $k<0$ and
$\alpha>-1/3$, the universe is accelerating for $\rho_b>\rho_c$ (i.e.
$a<a_c$) and decelerating for $\rho_b<\rho_c$ (i.e.  $a>a_c$).

In order to determine the temporal  evolution of $a(t)$, we shall assume that the universe is flat ($\kappa=0$).  Combining Eqs. (\ref{b7}) and (\ref{dark7}), we get
\begin{equation}
\label{dark10}
\dot a=\left \lbrack\frac{8\pi GA}{3}(1+\alpha)\right \rbrack^{1/2}\frac{a}{\sqrt{a^{3(1+\alpha)}-kA}}.
\end{equation}
For $a\rightarrow +\infty$, we recover the solution (\ref{dark3})-(\ref{dark4}). It is convenient to define $a_*=(|k|A)^{1/\lbrack 3(1+\alpha)\rbrack}$ and introduce $R=a/a_*$. In that case, the density is given by
\begin{equation}
\label{dark14}
\rho_b=\frac{1+\alpha}{|k|}\frac{1}{R^{3(1+\alpha)}\mp 1},
\end{equation}
and Eq. (\ref{dark10}) can be rewritten
\begin{equation}
\label{dark11}
\dot R=\frac{KR}{\sqrt{R^{3(1+\alpha)}\mp 1}},
\end{equation}
where the upper sign $-$ corresponds to $k>0$ and the lower sign $+$ corresponds to $k<0$. On the other hand, we have defined the constant
\begin{equation}
\label{dark12}
K=\left \lbrack \frac{8\pi G}{3|k|}(1+\alpha)\right \rbrack^{1/2}.
\end{equation}

\begin{figure}[!h]
\begin{center}
\includegraphics[clip,scale=0.3]{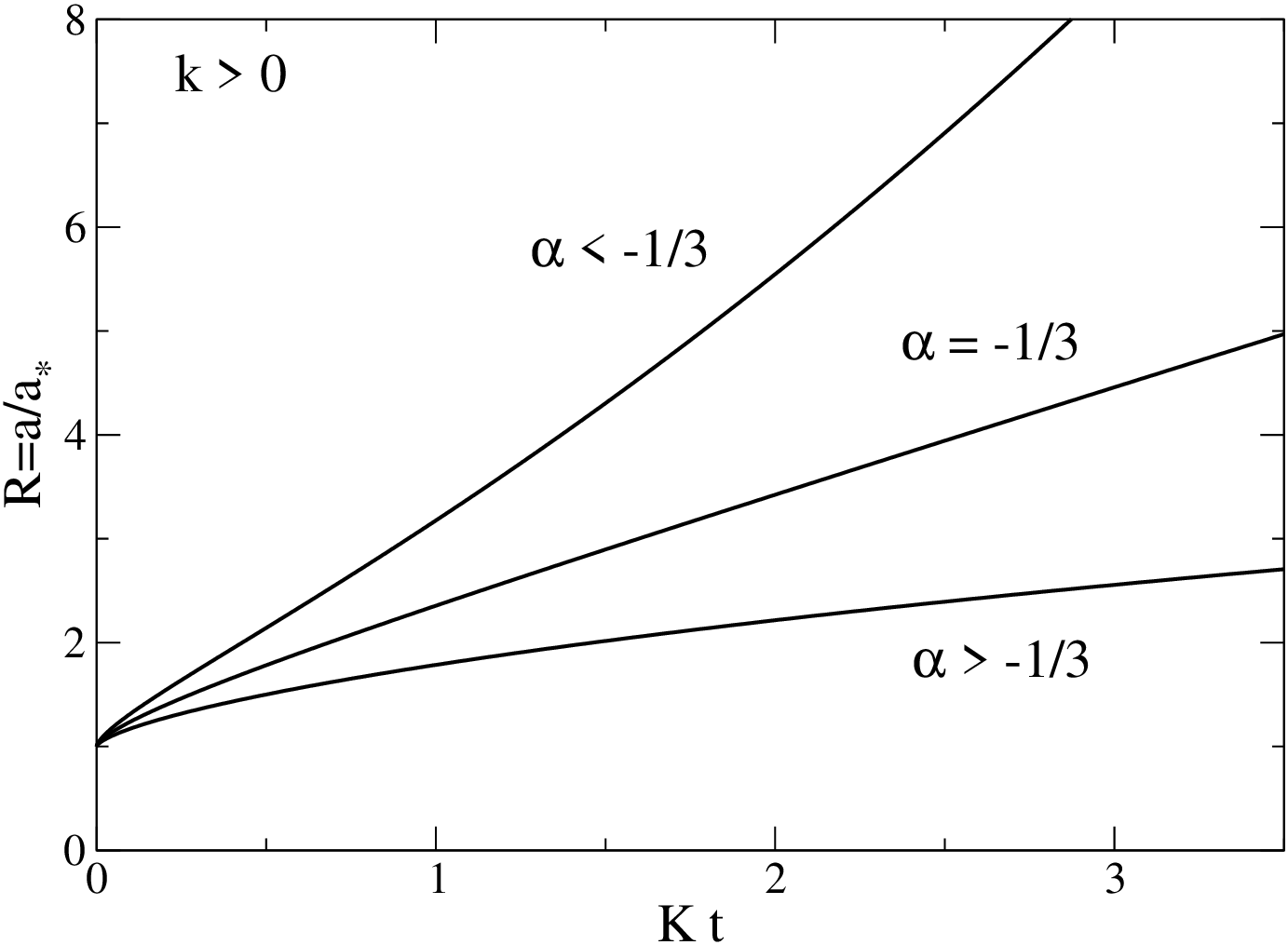}
\caption{Evolution of the scale factor in the case $k>0$ for different values of $\alpha$ (specifically $\alpha=-2/3$, $\alpha=-1/3$ and $\alpha=1/2$).}
\label{generalisationPOS}
\end{center}
\end{figure}

For $k>0$, the solution of Eq. (\ref{dark11}) is
\begin{equation}
\label{dark16}
\sqrt{R^{3(1+\alpha)}-1}-\arctan \sqrt{R^{3(1+\alpha)}-1}=\frac{3(1+\alpha)}{2}Kt,
\end{equation}
where the constant of integration has been set equal to zero. In this model, the universe starts at a finite time $t=0$ (see Sec. \ref{sec_gtlp} for a revision of this statement) with a finite radius $R(0)=1$ and an infinite density $\rho_b(0)=\infty$. For $t\rightarrow 0$, $R\simeq 1+\lbrack 3/(4(1+\alpha))\rbrack^{1/3}(Kt)^{2/3}$. For $t\rightarrow +\infty$, it asymptotically approaches the solution (\ref{dark3})-(\ref{dark4}) i.e.  $R\sim (3(1+\alpha)Kt/2)^{2/\lbrack 3(1+\alpha)\rbrack}$. When $\alpha\ge -1/3$, the universe is always decelerating ($\ddot R<0$). When $\alpha<-1/3$, the universe is decelerating for $\rho_b>\rho_c$ (i.e. $R<R_c^{+}\equiv ({-2}/(1+3\alpha))^{{1}/\lbrack {3(1+\alpha)}\rbrack}$) and accelerating for $\rho_b<\rho_c$ (i.e. $R>R_c^+$). The time at which the universe starts accelerating is
\begin{equation}
\label{dark17}
\frac{3(1+\alpha)}{2}Kt_c=\sqrt{\frac{-3(1+\alpha)}{1+3\alpha}}-\arctan \sqrt{\frac{-3(1+\alpha)}{1+3\alpha}}.
\end{equation}
Some possible evolutions of $R(t)$ corresponding to $\alpha<-1/3$, $\alpha=1/3$ and $\alpha>1/3$ are represented in Fig.  \ref{generalisationPOS}.

\begin{figure}[!h]
\begin{center}
\includegraphics[clip,scale=0.3]{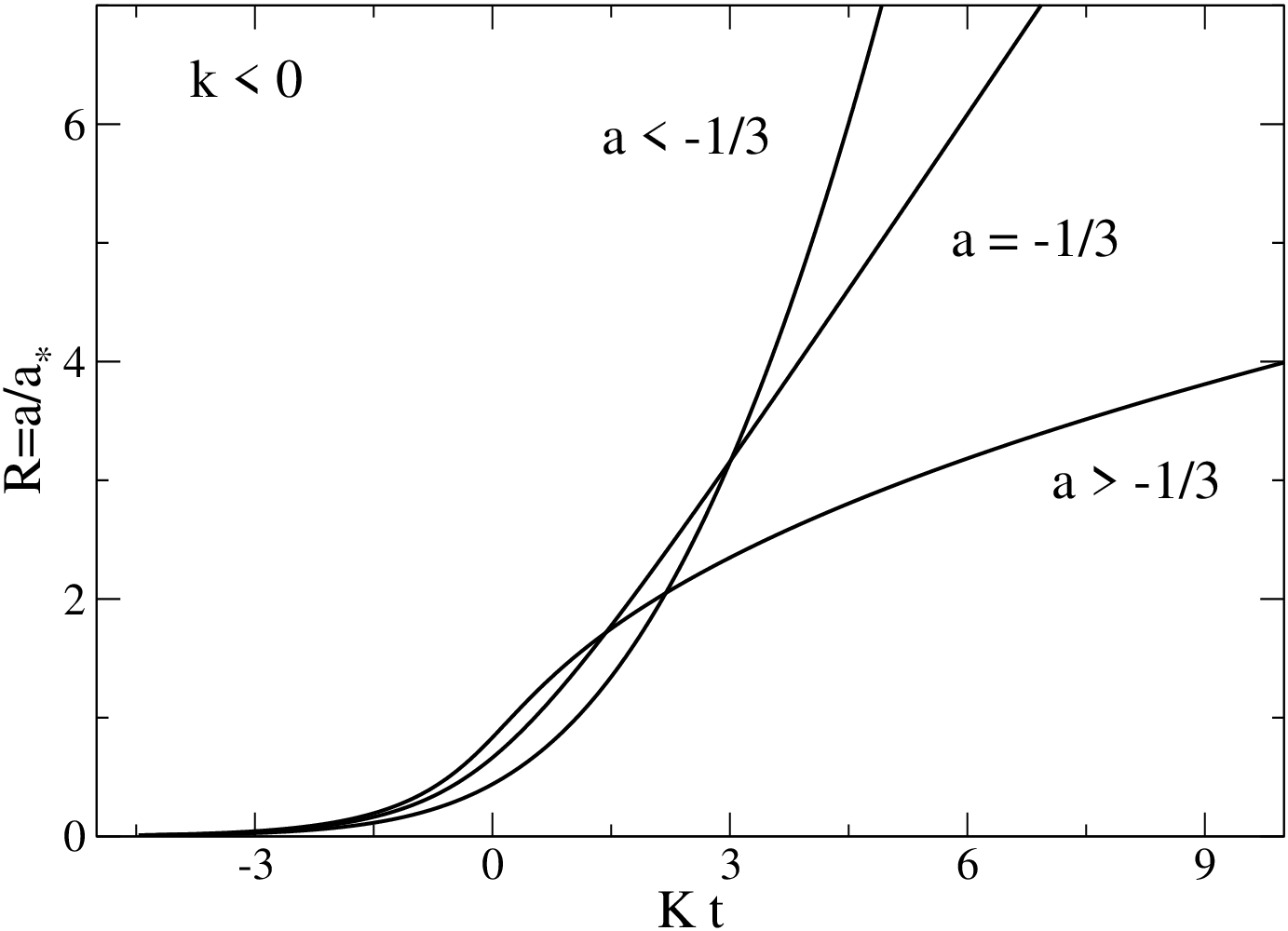}
\caption{Evolution of the scale factor in the case $k<0$ for different values of $\alpha$ (specifically $\alpha=-2/3$, $\alpha=-1/3$ and $\alpha=1/2$).}
\label{generalisationNEG}
\end{center}
\end{figure}

For $k<0$, the solution of Eq. (\ref{dark11}) is
\begin{equation}
\label{dark18}
\sqrt{R^{3(1+\alpha)}+1}-\ln \left (\frac{1+\sqrt{R^{3(1+\alpha)}+1}}{R^{3(1+\alpha)/2}}\right )=\frac{3(1+\alpha)}{2}Kt,
\end{equation}
where the constant of integration has been set equal to zero. In this model, the universe starts from  $t\rightarrow -\infty$ with a vanishing radius $R(-\infty)=0$ and a finite density $\rho_b(-\infty)=(1+\alpha)/|k|$. For $t\rightarrow +\infty$, it asymptotically approaches the solution (\ref{dark3})-(\ref{dark4}) i.e.  $R\sim (3(1+\alpha)Kt/2)^{2/\lbrack 3(1+\alpha)\rbrack}$.  When $\alpha\le -1/3$, the universe is always accelerating ($\ddot a>0$). When $\alpha>-1/3$, the universe is accelerating for $\rho_b>\rho_c$ (i.e. $R<R_c^{-}\equiv ({2}/(1+3\alpha))^{{1}/\lbrack {3(1+\alpha)}\rbrack}$) and decelerating for $\rho_b<\rho_c$ (i.e. $a>a_c$).  The time at which the universe starts decelerating is
\begin{equation}
\label{dark19}
\frac{3(1+\alpha)}{2}Kt_c=\sqrt{\frac{3(1+\alpha)}{1+3\alpha}}-\ln \left (\frac{\sqrt{1+3\alpha}+\sqrt{3(1+\alpha)}}{\sqrt{2}}\right ).
\end{equation}
For $t\rightarrow -\infty$,
$R\rightarrow 0$ and the asymptotic expansion of
Eq. (\ref{dark18}) yields
\begin{equation}
\label{dark20}
R(t)\propto e^{Kt}.
\end{equation}
Some possible evolutions of $R(t)$ corresponding to $\alpha<-1/3$, $\alpha=1/3$ and $\alpha>1/3$ are represented in Fig. \ref{generalisationNEG}.

\subsection{Optimal parameters}
\label{sec_op}

We can easily extend the analysis of Sec. \ref{sec_rep} to the equation of state (\ref{dark5}). The density can be written
\begin{eqnarray}
\label{dark21}
\rho_b=\rho_0\frac{1-q}{(\frac{a}{a_0})^{3(1+\alpha)}-q},
\end{eqnarray}
where
\begin{eqnarray}
\label{dark22}
q=\frac{k\rho_0}{1+\alpha+k\rho_0}, \qquad {\rm implying} \qquad k\rho_0=\frac{q(1+\alpha)}{1-q}.
\end{eqnarray}
We note that $q$ is positive for $k>0$ and negative for $k<0$. Furthermore, $q=0$ for $k=0$,  $q\rightarrow 1$ for $k\rightarrow +\infty$ and $q\rightarrow -\infty$ for $k\rightarrow -(1+\alpha)/\rho_0$. Using these results, the differential equation (\ref{b7}) giving the evolution of the scale factor with the time can be rewritten
\begin{eqnarray}
\label{dark23}
\dot a=H_0\sqrt{\Omega_0}(1-q)^{1/2}\frac{a}{\sqrt{(\frac{a}{a_0})^{3(1+\alpha)}-q}}.
\end{eqnarray}
We can make the connection with the previous notations by setting
\begin{eqnarray}
\label{dark24}
a_*=a_0 |q|^{1/\lbrack 3(1+\alpha)\rbrack},\qquad K=H_0\sqrt{\Omega_0}\left (\frac{1-q}{|q|}\right )^{1/2}.\qquad
\end{eqnarray}
In that case, Eq. (\ref{dark23}) takes the form of Eq. (\ref{dark11}) and its solutions are given by Eqs. (\ref{dark16}) and (\ref{dark18}). Returning to the notations of this section, we find that the evolution of $a(t)$ is given by Eqs. (\ref{rep9})-(\ref{rep11}) where all the  $3$ are replaced by $3(1+\alpha)$. The other equations can be easily generalized.

Let us assume that all the content of the universe is in the dark fluid with the generalized equation of state (\ref{dark5}) so that $\Omega_0=1$. Let us call $t_0$ the time at which $a/a_0=1$. For $0<q<1$,
\begin{eqnarray}
\label{dark25}
H_0 t_0=\frac{2}{3(1+\alpha)}\left\lbrace   1-\sqrt{\frac{q}{1-q}}\arctan \sqrt{\frac{1-q}{q}}\right\rbrace,
\end{eqnarray}
for $q<0$,
\begin{eqnarray}
\label{dark26}
H_0 t_0=\frac{2}{3(1+\alpha)}\left\lbrace 1-\sqrt{\frac{-q}{1-q}}
\ln(\sqrt{-q}+\sqrt{1-q})\right\rbrace,\nonumber\\
\end{eqnarray}
and for $q=0$,
\begin{eqnarray}
\label{dark27}
H_0 t_0=\frac{2}{3(1+\alpha)}.
\end{eqnarray}
These expressions basically give the age of the universe $t_0$ as a function of $q$ and $\alpha$.

\begin{figure}[!h]
\begin{center}
\includegraphics[clip,scale=0.3]{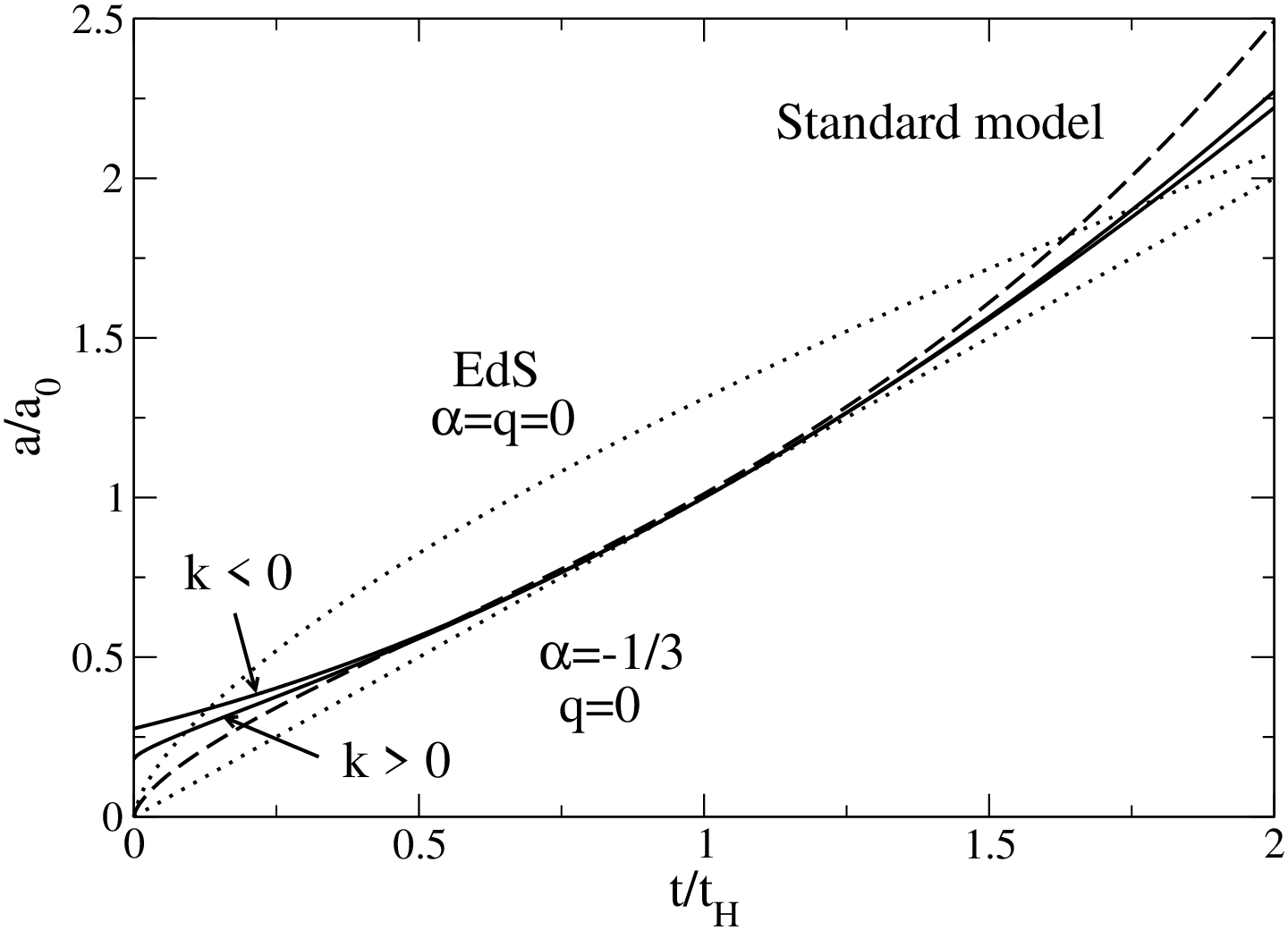}
\caption{The full line represents the scale factor $a(t)$ as a function of time $t$ for a universe filled with a dark fluid with equation of state (\ref{dark5}) with optimal parameters ($q=0.303471$, $\alpha=-0.76782367$) and ($q=-0.383589$, $\alpha=-0.53880367$). The long-dashed line corresponds to the standard model (see Figs. \ref{darkPOS} and \ref{darkNEG}). By construction, these three curves give the same values of the age of the universe $t_0\sim H_0^{-1}$ and of the present-day deceleration parameter $Q_0\sim -0.5$. The model with $k>0$ is decelerating then accelerating. Using Eqs. (\ref{dark17}) and (\ref{dark24}), we find that the acceleration starts at $H_0t_c\sim 0.2$ which is earlier than the value $H_0t_c\sim 0.5$ predicted by the standard model.  The model with $k<0$ is always accelerating.   We have also indicated the model corresponding to a linear equation of state ($q=0$) with $\alpha=-1/3$ yielding $a/a_0=H_0t$. It gives the correct age of the universe but has a vanishing deceleration parameter. Finally, we have indicated the (pressureless) EdS model corresponding to $\alpha=q=0$ yielding $a/a_0=(3H_0t/2)^{2/3}$. It does not give the correct age of the universe and is decelerating instead of accelerating. The other models with $\alpha=0$ and $q\neq 0$ do not do better. This shows that we must combine the two equations of state $\alpha\rho$ and $k\rho^2$ to obtain a good agreement with the standard model, i.e. we need an equation of state with two parameters. }
\label{modeleoptimal}
\end{center}
\end{figure}

The deceleration parameter is defined by
\begin{eqnarray}
\label{dark28}
Q=-\frac{{\ddot a} a}{{\dot a}^2}=\frac{1+3w(t)}{2},
\end{eqnarray}
where $w(t)\equiv p_b/\rho_b c^2$. For the equation of state (\ref{dark5}), $w(t)=\alpha+k\rho_b$. Using Eq. (\ref{dark22}), we find that the present-day value of the deceleration parameter can be expressed in our model as
\begin{eqnarray}
\label{dark29}
Q_0=\frac{3\alpha+2q+1}{2(1-q)}.
\end{eqnarray}
The observed value of $Q_0$ is close to $-0.5$. If we take $Q_0=-1/2$, Eq. (\ref{dark29}) yields $\alpha=-(q+2)/3$. On the other hand, in the standard model, the age of the universe is about $13.75\, {\rm Gyrs}$. If we take  $t_0=H_0^{-1}$ (which is close to this value) and use the previous relation between $\alpha$ and $q$, Eqs. (\ref{dark25}) and (\ref{dark26}) determine the optimal value of $q$, then $\alpha$.  For $0<q<1$, the optimal value of $q$ is given by
\begin{eqnarray}
\label{dark30}
\frac{1+q}{2}-\sqrt{\frac{q}{1-q}}\arctan \sqrt{\frac{1-q}{q}}=0.
\end{eqnarray}
We find $q=0.303471$ and $\alpha=-0.76782367$. For $q<0$, the optimal value of $q$ is given by
\begin{eqnarray}
\label{dark31}
\frac{1+q}{2}-\sqrt{\frac{-q}{1-q}}\ln(\sqrt{-q}+\sqrt{1-q}) =0.
\end{eqnarray}
We find  $q=-0.383589$ and $\alpha=-0.53880367$. The evolution of the scale factor $a(t)$ in a universe filled with a dark fluid with equation of state (\ref{dark5}) is represented in Fig. \ref{modeleoptimal} for the optimal values of $(q,\alpha)$ obtained previously. It gives a relatively good agreement with the standard $\Lambda$CDM model based on a non-vanishing value of the cosmological constant.

\subsection{Theory of linear perturbations}
\label{sec_gtlp}

In this section, we extend to the case of the generalized equation of
state (\ref{dark5}) the theory of linear perturbations developed in
Sec. \ref{sec_tl}. For this equation of state, $w=\alpha+k\rho_b$ and
$c_{eff}^2=c_s^2=2w-\alpha$. According to
Eq. (\ref{dark14}), we have
\begin{eqnarray}
\label{dark32}
w=\alpha\pm\frac{1+\alpha}{R^{3(1+\alpha)}\mp 1},
\end{eqnarray}
where the upper sign corresponds to $k>0$ and the lower sign corresponds to $k<0$. The generalization of Eq. (\ref{tl6}) is
\begin{eqnarray}
\label{dark33}
R^2\frac{d^2 \delta}{dR^2}+\frac{3}{2}(1+3w-4\alpha) R \frac{d\delta}{dR}\nonumber\\
-\frac{3}{2} (9w^2+12w+1-8\alpha-6\alpha^2)\delta=0.
\end{eqnarray}
For $R\rightarrow +\infty$, $w\rightarrow \alpha$ and Eq. (\ref{dark33}) reduces to
\begin{eqnarray}
\label{dark34}
R^2\frac{d^2 \delta}{dR^2}+\frac{3}{2}(1-\alpha) R \frac{d\delta}{dR}-\frac{3}{2}(\alpha+1)(3\alpha+1)\delta=0.\nonumber\\
\end{eqnarray}
This is the equation for the density contrast corresponding to the  linear equation of state (\ref{dark0}). Its solutions are
\begin{eqnarray}
\label{dark35}
\delta \propto R^{1+3\alpha},\qquad \delta \propto R^{-\frac{3}{2}(1+\alpha)}.
\end{eqnarray}

Let us now consider the case of small $R$. For $k>0$,  the requirement $c_{eff}=c_s<1$ implies $R>\lbrack(3+\alpha)/(1-\alpha)\rbrack^{1/\lbrack 3(1+\alpha)\rbrack}$ and $w<(\alpha+1)/2$. Therefore, we shall assume that the universe starts at $t_*$ such that $R(t_*)=\lbrack(3+\alpha)/(1-\alpha)\rbrack^{1/\lbrack 3(1+\alpha)\rbrack}$. According to Eq. (\ref{dark16}), we have
\begin{equation}
\label{dark36}
\frac{3(1+\alpha)}{2}Kt_*=\sqrt{\frac{2(1+\alpha)}{1-\alpha}}-\arctan \sqrt{\frac{2(1+\alpha)}{1-\alpha}}.
\end{equation}
For $t\rightarrow t_*$, Eq. (\ref{dark33}) can be approximated by
\begin{eqnarray}
\label{dark37}
R^2\frac{d^2 \delta}{dR^2}+\frac{15}{4}(1-\alpha) R \frac{d\delta}{dR}+\frac{3}{8}(15\alpha^2-10\alpha-37)\delta=0,\nonumber\\
\end{eqnarray}
and its solutions are
\begin{eqnarray}
\label{dark38}
\delta_{\pm} \propto R^{\frac{-11+15\alpha\pm\sqrt{-135\alpha^2-90\alpha+1009}}{8}}.
\end{eqnarray}
For $k<0$, the universe starts at $R=0$ for $t\rightarrow -\infty$ implying $w=-1$.
In that case,  Eq. (\ref{dark33}) can be approximated by
\begin{eqnarray}
\label{dark39}
R^2\frac{d^2 \delta}{dR^2}-3(1+2\alpha) R \frac{d\delta}{dR}+3(1+\alpha)(1+3\alpha)\delta=0,\quad
\end{eqnarray}
and its solutions are
\begin{eqnarray}
\label{dark40}
\delta \propto R^{3(1+\alpha)},\qquad \delta \propto R^{1+3\alpha}.
\end{eqnarray}

\section{Conclusion}
\label{sec_conclusion}

Following the proposal of B\"ohmer \& Harko (2007) and others (see a short historic in Chavanis 2011), we have assumed that the dark matter in the universe is a self-gravitating BEC with short-range interactions, and we have theoretically explored the consequences of this hypothesis. For the sake of generality, we have considered the case of positive and negative scattering lengths.

At the level of dark matter halos, a positive scattering length, equivalent to a repulsive self-interaction generating a positive pressure, is able to stabilize the halos with respect to gravitational collapse. This leads to dark matter halos without density cusps, equivalent to polytropes of index $n=1$ (more generally, the barotropic equation of state is fixed by the form of the self-interaction so that other configurations are possible). Alternatively, if the scattering length is negative, equivalent to an attractive self-interaction  generating a negative pressure, the dark matter is very unstable and collapses above a very small critical mass $M_{max}=1.012\hbar/\sqrt{|a_s|Gm}$ (Chavanis 2011). When these ideas are applied to an infinite homogeneous cosmic fluid (Jeans problem), it is found that a negative scattering length can increase the maximum growth rate of the instability and accelerate the formation of structures. The virtues of these results could be combined by assuming that the scattering length changes sign in the course of the evolution. It could be initially negative to help with the formation of structures and become positive (due to a change of density, magnetic fields, radiation,...) to prevent complete gravitational collapse. The mechanism of this change of sign is, however, unknown so that this idea remains highly speculative.  However, some terrestrial experiments have demonstrated that certain atoms can have negative scattering lengths, that their scattering length can depend on the magnetic field, and that it is possible in principle to manipulate the value and the sign of $a_s$ (Fedichev et al. 1996). Therefore, a BEC is a serious candidate for which the pressure can be positive and/or negative (see Appendix \ref{sec_det} for further remarks about the values of the BEC parameters).

At the cosmological level, we have constructed models of universe composed of BEC dark matter with attractive or repulsive self-interaction. We have first studied the academic situation where the universe is made only of BEC dark matter. A BEC universe with positive scattering length, having a positive pressure, is not qualitatively very different from a classical Einstein-de Sitter universe. It also emerges at a primordial time  $t=0$ from a big-bang singularity where the density is infinite, and undergoes a decelerating expansion asymptotically equivalent to the EdS universe. A difference, however, is that the initial scale factor $a(0)$ is finite.  On the other hand,  a BEC universe with negative scattering length, having a negative pressure,  markedly differs from previous models. It starts from $t\rightarrow -\infty$ with a vanishing radius and a finite density, has an initial accelerating expansion then decelerates and asymptotically behaves like the EdS universe. This model universe exists for any time in the past and there is no big-bang singularity. When we  add the effect of radiation, baryonic matter and dark energy (via the cosmological constant), the picture is different. In that case, a BEC universe with attractive or repulsive self-interaction starts from a singularity at $t=0$ where the density is infinite. It first experiences a phase of  decelerating expansion followed by a phase of accelerating expansion. For $k\rightarrow 0$ we recover the standard $\Lambda$CDM model but for $k\neq 0$, the evolution of the scale factor in a BEC universe can be substantially different. The model with  $k>0$ expands more rapidly than the standard model ($k=0$). The initial scale factor is finite ($a(0)>0$)  and the radiation never dominates. The model with $k<0$ expands less rapidly than the standard model. The initial scale factor vanishes  ($a(0)=0$) and the  radiation dominates leading to a decelerating expansion. In both models, the dark energy dominates at large times leading to an accelerating expansion.  Finally, we have  considered a ``dark fluid'' with generalized equation of state $p=(\alpha\rho+k\rho^2)c^2$ having a component $p=k\rho^2 c^2$ similar to a BEC dark matter and a component $p=\alpha\rho c^2$ mimicking the effect of the cosmological constant (dark energy). We have found optimal parameters $(\alpha,k)$ that give a good agreement with the standard model. We have studied the growth of perturbations in these different models and confirmed the previous observation of Harko (2011) that the density contrast increases more rapidly in a BEC universe than in the standard model. 

In conclusion, the idea that dark matter could be a BEC is  fascinating and probably deserves further research.

\appendix

\section{Solution of the differential equation (\ref{eds5})}
\label{sec_sol}

In this Appendix, we provide some analytical solutions of the differential equation (\ref{eds5}). If we define
\begin{eqnarray}
\label{sol1}
\mu=\frac{3k^4}{2\kappa_Q^4},\qquad \lambda=\frac{3k^2}{2\kappa_J^2}, \qquad \alpha=\frac{4-3\gamma}{2},
\end{eqnarray}
this equation can be written
\begin{eqnarray}
\label{sol2}
\frac{d^2 f}{da^2}-\frac{1}{2a}\frac{df}{da}+\frac{1}{a^2}\left (\frac{\mu}{a}+\lambda a^{2\alpha}-1\right )f=0.
\end{eqnarray}
With the change of function $f(a)=a^{3/4}g(a)$, we get
\begin{eqnarray}
\label{sol3}
\frac{d^2 g}{da^2}+\frac{1}{a}\frac{dg}{da}+\frac{1}{a^2}\left (\frac{\mu}{a}+\lambda a^{2\alpha}-\frac{25}{16}\right )g=0.
\end{eqnarray}
Assuming $\alpha\neq 0$ and defining $x=a^{\alpha}$, we obtain
\begin{eqnarray}
\label{sol4}
x^2\frac{d^2 g}{dx^2}+x\frac{dg}{dx}+\frac{1}{\alpha^2}\left (\frac{\mu}{x^{1/\alpha}}+\lambda x^2-\frac{25}{16}\right )g=0.
\end{eqnarray}
We have not been able to find the general solution of this differential equation in terms of simple functions. We therefore consider particular cases. We first consider the case $\mu=0$. Then
\begin{eqnarray}
\label{sol5}
x^2\frac{d^2 g}{dx^2}+x\frac{dg}{dx}+\frac{1}{\alpha^2}\left (\lambda x^2-\frac{25}{16}\right )g=0.
\end{eqnarray}
For $\lambda>0$, defining $y=\sqrt{\lambda}x/\alpha$ and $p=5/4\alpha$, the foregoing equation becomes
\begin{eqnarray}
\label{sol6}
y^2\frac{d^2 g}{dy^2}+y\frac{dg}{dy}+\left (y^2-p^2\right )g=0.
\end{eqnarray}
This is a Bessel equation whose regular solutions are $g=J_{\pm p}(y)$. Coming back to the original variables, we get
\begin{eqnarray}
\label{sol7}
f(a)\propto a^{3/4}J_{\pm \frac{5}{4\alpha}}\left (\frac{\sqrt{\lambda}a^{\alpha}}{\alpha}\right ).
\end{eqnarray}
For $\lambda<0$, defining $y=\sqrt{-\lambda}x/\alpha$ and $p=5/4\alpha$, Eq. (\ref{sol5}) becomes
\begin{eqnarray}
\label{sol8}
y^2\frac{d^2 g}{dy^2}+y\frac{dg}{dy}-\left (y^2+p^2\right )g=0.
\end{eqnarray}
This is a Bessel equation whose regular solutions are $g=I_{\pm p}(y)$. Coming back to the original variables, we get
\begin{eqnarray}
\label{sol9}
f(a)\propto a^{3/4}I_{\pm \frac{5}{4\alpha}}\left (\frac{\sqrt{-\lambda}a^{\alpha}}{\alpha}\right ).
\end{eqnarray}

For  $\alpha=0$ and $\mu=0$, Eq. (\ref{sol2}) becomes
\begin{eqnarray}
\label{sol10}
\frac{d^2 f}{da^2}-\frac{1}{2a}\frac{df}{da}+\frac{1}{a^2}\left (\lambda-1\right )f=0.
\end{eqnarray}
Looking for solutions of the form $f(a)\propto a^n$ we find that
\begin{eqnarray}
\label{sol11}
f(a)\propto a^{\frac{3}{4}\pm \sqrt{\frac{25}{16}-\lambda}}.
\end{eqnarray}

Finally, for  $\alpha=0$ and $\mu\neq 0$, Eq. (\ref{sol2}) can be rewritten
\begin{eqnarray}
\label{sol12}
\frac{d^2 f}{da^2}-\frac{1}{2a}\frac{df}{da}+\frac{1}{a^2}\left (\frac{\mu}{a}+\lambda-1\right )f=0.
\end{eqnarray}
Proceeding as before, we find that the solution of this equation is
\begin{eqnarray}
\label{sol13}
f(a)\propto a^{3/4}J_{\pm \frac{1}{2}\sqrt{25-16\lambda}}\left (2\sqrt{\frac{\mu}{a}}\right ).
\end{eqnarray}

\section{Estimate of the BEC parameters}
\label{sec_det}

In this Appendix, we derive some constraints on the BEC parameters. The parameter $k$ appearing in the equation of state (\ref{b4})  of a BEC is defined by $k=2\pi a_s\hbar^2/m^3c^2$ and the present-day critical density is $(\rho_c)_0=3H_0^2/8\pi G$. Introducing the dimensionless parameter $\lambda=a_s mc/\hbar$ (Chavanis 2011), we obtain the general relations
\begin{eqnarray}
\label{det1}
k(\rho_c)_0&=&\frac{3}{4}(H_0t_p)^2\left (\frac{M_p}{m}\right )^4\lambda\nonumber\\
&=&2.457\, 10^{-10}\left (\frac{{\rm eV}/c^2}{m}\right )^4\lambda,
\end{eqnarray}
and
\begin{eqnarray}
\label{det2}
\lambda=5.091\, 10^{-9} \frac{a_s}{\rm fm}\frac{m}{{\rm eV}/c^2}.
\end{eqnarray}
On the other hand, the radius of a BEC dark matter halo is given by (Arbey et al. 2003, B\"ohmer \& Harko 2007, Chavanis 2011):
\begin{eqnarray}
\label{det3}
R=\pi\left (\frac{a_s \hbar^2}{Gm^3}\right )^{1/2}=\pi\left (\frac{\lambda\hbar^3}{Gc}\right )^{1/2}\frac{1}{m^2},
\end{eqnarray}
so that
\begin{eqnarray}
\label{det4}
\left (\frac{R}{\rm kpc}\right )^2=5.951\, 10^4 \left (\frac{{\rm eV}/c^2}{m}\right )^4\lambda.
\end{eqnarray}
{\it Therefore, the typical size of the dark matter halos determines the ratio $\lambda/m^4$. }
Taking $R=10\, {\rm kpc}$, we obtain $(m/({\rm eV}/c^2))/\lambda^{1/4}=4.94$. Now, we remark that the parameter  $k(\rho_c)_0$, given by Eq. (\ref{det1}), only depends on  $\lambda/m^4$. Using the previous estimate, we get $k(\rho_c)_0=4.126\, 10^{-13}$. We conclude that the dimensionless parameter $q$, defined by Eq. (\ref{rep3}), is very small.

This estimate, which does not rely on any free parameter, seems to indicate that the pressure of the BEC dark matter is totally negligible at the scale of the cosmic fluid (since $q\simeq 0$ like in a pressureless universe) while it is important at the scale of galactic halos (since $R^2\sim \lambda/m^4$). This seems to be bad news for the BEC cosmology. In fact, this result implies either that (i) the BEC dark matter can be treated as a pressureless fluid, like the ordinary dark matter, at the cosmological scale (i.e. in the Friedmann equations) or that (ii) the BEC parameters (in particular the scattering length $a_s$) are different in the homogeneous cosmic fluid (before the Jeans instability) and in the dark matter halos (after the Jeans instability). This observation may be a hint that the value of the scattering length changes in the course of time and that the scattering length of the bosons in the dark halos is not the same as in the cosmic fluid (because the density is different). This corroborates our remark (Chavanis 2011) that the sign of the scattering length may change too.  Clearly, the determination of the BEC parameters, or the constraints that they must satisfy, is certainly a very important step for the validation of the BEC dark matter hypothesis.

\section{Equation of state $p=(-\rho+k\rho^2)c^2$}
\label{sec_eosg}

In this Appendix, we consider the equation of state (\ref{dark5}) with $\alpha=-1$ and $k\neq 0$. In that case, Eq. (\ref{dark6}) can be integrated into
\begin{eqnarray}
\label{eosg1}
\rho_b=\frac{1}{3k\ln(a/a_*)},
\end{eqnarray}
where $a_*$ is a constant. Physical solutions  require that $k>0$ and $a\ge a_*$. Setting $R=a/a_*$, the Friedmann equation (\ref{b7}) with $\kappa=\Lambda=0$ can be written
\begin{eqnarray}
\label{eosg2}
\dot R=\frac{2}{3}K\frac{R}{\sqrt{\ln R}},
\end{eqnarray}
where $K=(2\pi G/k)^{1/2}$. The solution of Eq. (\ref{eosg2}) is
\begin{eqnarray}
\label{eosg3}
R(t)=e^{(Kt)^{2/3}}.
\end{eqnarray}
For $t\rightarrow 0$, $R\simeq 1+(Kt)^{2/3}$. The curve $R(t)$
presents an inflexion point at $R_c=\sqrt{e}$, $(\rho_b)_c=2/(3k)$ and
$Kt_c=(1/2)^{3/2}$. The expansion is decelerating for $t<t_c$ and
accelerating for $t>t_c$.

If we introduce the present-day values $\rho_0$ and $a_0$ of the density and scale factor, the density can be rewritten
\begin{eqnarray}
\label{eosg4}
\rho_b=\frac{\rho_0}{3q\ln(a/a_0)+1},
\end{eqnarray}
where $q=k\rho_0$. We have $a_*/a_0={\rm exp}(-1/3q)$ and $K=(3\Omega_0/4q)^{1/2}H_0$. Using Eq. (\ref{eosg3}) the evolution of the scale factor is given by
\begin{eqnarray}
\label{eosg5}
\frac{a(t)}{a_0}=e^{-\frac{1}{3q}}e^{(\frac{3\Omega_0}{4q})^{1/3}(H_0t)^{2/3}}.
\end{eqnarray}
The time at which the universe starts accelerating is $H_0t_c=(q/6\Omega_0)^{1/2}$. The age of the universe, corresponding to the time $t=t_0$ at which $a/a_0=1$ is $H_0t_0=2/(9q)$ (we have taken $\Omega_0=1$). Finally, the present-day value of the deceleration parameter (\ref{dark28}) is $Q_0=(3q-2)/2$.


\begin{thebibliography}{}


\bibitem{arbey}{\small Arbey A., Lesgourgues J.,  Salati P., 2003,
Phys. Rev. D {68}, 023511}
\bibitem{baldeschi}{\small Baldeschi M.R.,  Gelmini G.B., Ruffini R., 1983, Phys. Lett. B {122}, 221}
\bibitem{bialynicki}{\small Bialynicki-Birula I., Mycielski J., 1976, Ann. Phys. {100}, 62}
\bibitem{bianchi}{\small Bianchi M., Grasso D., Ruffini R., 1990, A\&A {231}, 301}
\bibitem{bilic}{\small Bilic N.,  Lindebaum R.J., Tupper G.B., Viollier R.D., 2001, Phys. Lett. B {515}, 105}
\bibitem{bt}{\small Binney J., Tremaine S., 1987, Galactic Dynamics.  Princeton University Press}
\bibitem{bohmer}{\small B\"ohmer C.G., Harko T., 2007, J. Cosmol. Astropart. Phys. 06, 025}
\bibitem{bonnor}{\small Bonnor W.B., 1957, MNRAS {117}, 104}
\bibitem{borriello}{\small Borriello A., Salucci P., 2001, MNRAS {323}, 285}
\bibitem{breit}{\small Breit J.D., Gupta S., Zaks A., 1984, Phys. Lett. B {140}, 329}
\bibitem{callan}{\small Callan C., Dicke R.H., Peebles P.J.E., 1965, Am. J. Phys.  {33}, 105}
\bibitem{chandra}{\small Chandrasekhar S., 1939, Stellar structure. University of Chicago Press}
\bibitem{cd}  {\small Chavanis P.H., Delfini L., 2010, Phys. Rev. E {81}, 051103}
\bibitem{paper1}{\small Chavanis P.H., 2011, e-print arXiv:1103.2050}
\bibitem{paper2}{\small Chavanis P.H., Delfini L., 2011, e-print arXiv:1103.2054}
\bibitem{colpi}{\small Colpi M., Shapiro S.L., Wasserman I., 1986, Phys. Rev. Lett. {57}, 2485}
\bibitem{revuebec}{\small Dalfovo F., Giorgini S., Pitaevskii L.P., Stringari S., 1999, Rev. Mod. Phys. {71}, 463}
\bibitem{novae3}{\small  de Bernardis P. {et al.}, 2000, Nature {404}, 995}
\bibitem{eddington}{\small Eddington A.S., 1930, MNRAS {90}, 668}
\bibitem{einstein}  {\small  Einstein A., 1917, Sitzungsber. Preuss. Akad. Wiss.  {1}, 142}
\bibitem{eds}{\small Einstein A., de Sitter W., 1932, Proc. Natl. Acad. Sci. (U.S.)  {18}, 213}
\bibitem{fedichev}{\small Fedichev P.O., Kagan Yu., Shlyapnikov G.V., Walraven J.T.M., 1996, Phys. Rev. Lett. {77}, 2913}
\bibitem{friedmann1}{\small Friedmann A., 1922, Z. Physik {10}, 377}
\bibitem{friedmann2}{\small Friedmann A., 1924, Z. Physik {21}, 326}
\bibitem{gilbert}{\small Gilbert I.H., 1966, ApJ  {144}, 233}
\bibitem{goodman}{\small Goodman J., 2000, New Astronomy {5}, 103}
\bibitem{guth}{\small Guth A.H., 1981, Phys. Rev. D {23}, 347}
\bibitem{novae4}{\small  Hanany S. {et al.}, 2000, ApJ {545}, L5}
\bibitem{harko}{\small Harko, T., 2011, to appear in MNRAS, e-print arXiv:1101.3655}
\bibitem{harri}{\small Harrison E.R., 1965, Ann. Phys. (N.Y.) {35}, 437}
\bibitem{harrison}{\small Harrison E.R., 1967, Rev. Mod. Phys. {39}, 862}
\bibitem{hinshaw}{\small Hinshaw G. et al., 2009, ApJ Supp. {180}, 225}
\bibitem{hu}{\small Hu W., Barkana R., Gruzinov A., 2000, Phys. Rev. Lett. {85}, 1158}
\bibitem{jeans1}{\small Jeans J.H., 1902, Phil. Trans. A {199}, 49}
\bibitem{jeans2}{\small Jeans J.H., 1929, Astronomy and Cosmogony.  Cambridge University Press}
\bibitem{kaup}{\small Kaup D.J., 1968,  Phys. Rev. {172}, 1331}
\bibitem{khlopov}{\small Khlopov M.Yu., Malomed B.A., Zeldovich Ya.B., 1985, MNRAS {215}, 575}
\bibitem{kiessling}{\small Kiessling M., 2003, Adv. Appl. Math. {31}, 132; see also e-print arXiv:astro-ph/9910247}
\bibitem{layzer}{\small Layzer D., 1954, Astron. J. {59}, 268}
\bibitem{leekoh}{\small Lee J., Koh I., 1996, Phys. Rev. D {53}, 2236}
\bibitem{lima}{\small Lima J.A.S., Zanchin V., Brandenberger R., 1997, MNRAS {291}, L1}
\bibitem{madelung}{\small Madelung E., 1927, Zeit. F. Phys. {40}, 322}
\bibitem{milgrom}{\small Milgrom M., 1983, ApJ {270}, 365}
\bibitem{milne}{\small Milne E.A., 1934, Quarterly J. Math. {5}, 64}
\bibitem{mcCreamilne}{\small McCrea W.H., Milne E.A., 1934, Quarterly J. Math. {5}, 73}
\bibitem{mcCrea}{\small McCrea W.H., 1951,  Proc. R. Soc. London. {206}, 562}
\bibitem{mC}{\small McCrea W.H., 1955, Astron. J. {60}, 271}
\bibitem{overduin}{\small Overduin J.M., Wesson P.S., 2004, Phys. Rep.  {402}, 267}
\bibitem{pace}{\small Pace F., Waizmann J.C., Bartelmann M., 2010, MNRAS {406}, 1865}
\bibitem{peeblesbook}{\small Peebles P.J.E., 1980, The Large-Scale Structure of the Universe. Princeton University Press}
\bibitem{peebles}{\small Peebles P.J.E., 2000, ApJ {534}, L127}
\bibitem{peebleslambda}{\small Peebles P.J.E., Ratra B., 2003, Rev. Mod. Phys. {75} 559}
\bibitem{novae2}{\small Perlmutter S. {et al.}, 1999, ApJ {517}, 565}
\bibitem{primack}{\small Primack J.R., Seckel D., Sadoulet B., 1988, Annu. Rev. Nucl. Part. Sci.  {38}, 75}
\bibitem{reis}{\small Reis R.R.R., 2003, Phys. Rev. D {67}, 087301; Erratum: Phys. Rev. D {68}, 089901(E)}
\bibitem{novae1}{\small Riess A.G. {et al.}, 1998, Astron. J. {116}, 109}
\bibitem{rb}{\small Ruffini R., Bonazzola S., 1969, Phys. Rev. {187}, 1767}
\bibitem{savedoff}{\small Savedoff M.P., Vila S., 1962, ApJ {136}, 609}
\bibitem{sin}{\small Sin S.J., 1994, Phys. Rev. D {50}, 3650}
\bibitem{sikivie}{\small Sikivie P., Yang Q., 2009, Phys. Rev. Lett. {103}, 111301}
\bibitem{spiegel}  {\small  Spiegel E.A., 1998,  in {Gravitational Screening}, edited by A. Harvey (Springer-Verlag, Heidelberg); e-print arXiv:astro-ph/9801014}
\bibitem{sulem}{\small Sulem C., Sulem P.L., 1999, The Nonlinear Schr\"odinger Equation. Springer}
\bibitem{takasugi}{\small Takasugi E., Yoshimura M., 1984, Z.  Phys. C {26}, 241}
\bibitem{thirring}{\small Thirring W., 1983,  Phys. Lett. B {127}, 27}
\bibitem{bij}{\small van der Bij J.J., Gleiser M., 1987,  Phys. Lett. B {194}, 482}
\bibitem{weinberg}{\small Weinberg S., 1972, Gravitation and Cosmology. John Wiley \& Sons}
\bibitem{wk}{\small Widrow L.M., Kaiser N., 1993, ApJ {416}, L71}
\bibitem{zwicky}{\small Zwicky F., 1937,  ApJ {86}, 217}



\end{thebibliography}
\end{document}